\documentclass[acmlarge, nonacm]{acmart}
\AtBeginDocument{%
  }
\setcopyright{acmlicensed}

\usepackage{listings}
\usepackage{xcolor}
\usepackage{textcomp} 
\usepackage{booktabs}
\usepackage{tabularx}
\usepackage{tikz}
\usetikzlibrary{arrows.meta, positioning, fit, backgrounds, calc}

\definecolor{lightgray}{rgb}{0.95, 0.95, 0.92}
\definecolor{darkgray}{rgb}{0.4, 0.4, 0.4}
\definecolor{purple}{rgb}{0.65, 0.12, 0.82}
\definecolor{blue}{rgb}{0.13, 0.13, 1}
\definecolor{red}{rgb}{0.9, 0, 0}
\definecolor{green}{rgb}{0, 0.2, 0}
\definecolor{placeholder}{rgb}{0.8, 0.4, 0.0} 

\lstdefinelanguage{JavaScript}{
  keywords={typeof, new, true, false, catch, function, return, null, catch, switch, var, if, in, of, for, continue, while, do, else, case, break, let, const, async, await},
  keywordstyle=\color{blue}\bfseries,
  ndkeywords={class, export, boolean, throw, implements, import, this},
  ndkeywordstyle=\color{darkgray}\bfseries,
  identifierstyle=\color{black},
  sensitive=false,
  comment=[l]{//},
  morecomment=[s]{/*}{*/},
  commentstyle=\color{purple}\ttfamily,
  stringstyle=\color{green}\ttfamily,
  morestring=[b]',
  morestring=[b]",
  morestring=[b]`,
  moredelim=*[s][\color{placeholder}\bfseries]{\$\{}{\}}
}

\tikzset{
  box/.style={draw, rounded corners=2pt, align=center, minimum height=0.95cm, text width=2.5cm, font=\footnotesize, fill=white},
  llm/.style={box, fill=orange!12},
  local/.style={box, fill=green!8},
  med/.style={box, fill=blue!8},
  arr/.style={-{Stealth[length=2mm]}, thick},
  darr/.style={-{Stealth[length=2mm]}, thick, dashed},
  lbl/.style={font=\scriptsize\itshape, align=center}
}

\newcommand{\fold}[2][0.22]{%
  \fill[white] ($(#2.north east)+(-#1,0)$) -- (#2.north east) -- ($(#2.north east)+(0,-#1)$) -- cycle;
  \draw[fill=gray!32] ($(#2.north east)+(-#1,0)$) -- ($(#2.north east)+(0,-#1)$) -- ($(#2.north east)+(-#1,-#1)$) -- cycle;
}

\begin{document}

\title{Metadata-Driven AI: An Architectural Pattern for Sustainable Data Analysis with LLMs}

\author{Renaud Pawlak}
\affiliation{%
  \institution{Luxembourg Institute of Science and Technology (LIST)}
  \city{Esch-sur-Alzette}
  \country{Luxembourg}
}
\email{renaud.pawlak@list.lu}

\author{Armen Sulejmani}
\affiliation{%
  \institution{Luxembourg Institute of Science and Technology (LIST)}
  \city{Esch-sur-Alzette}
  \country{Luxembourg}
}
\email{armen.sulejmani@list.lu}

\author{Jordi Cabot}
\affiliation{%
  \institution{Luxembourg Institute of Science and Technology (LIST)}
  \city{Esch-sur-Alzette}
  \country{Luxembourg}
}
\email{jordi.cabot@list.lu}

\begin{abstract}
Large Language Models (LLMs) let users analyse data in natural language, either by giving the data to the model to read directly or by having the model generate code to process it. Both routes send the data, or fragments of it, to the model; both pay for inference on every run; and the code a model writes to answer a question is usually discarded once the question is answered. We present \textit{Metadata-Driven AI}, an architectural pattern that keeps the data away from the model and turns each analysis into a reusable program. Only \textit{metadata} (not the actual data) crosses a one-way \textit{inference boundary} to the LLM. 
The program the LLM returns runs next to the data and is kept, so that inference is paid once per analysis rather than once per run. We describe the pattern together with a simple cost model, a measurement of how little information crosses the boundary on public datasets, and a benchmark showing that a small, locally hosted model can serve as the generator. We illustrate the pattern on three applications, each built on a different implementation of it: geospatial photovoltaic prospection over live public APIs, extraction of portfolio positions from bank statements, and a conversational assistant over smart-city dashboards.
\end{abstract}


\maketitle

\section{Introduction}
\label{sec:intro}

Large Language Models (LLMs) have changed how analysts interact with data: instead of writing queries, scripts, or spreadsheet formulas, users can describe what they want in natural language \cite{li2023autonomous, janowicz2020geoai}. A model can satisfy such a request in two ways. It can read the data into its context window --- now large enough to hold sizeable files --- and produce the answer directly by inference. Or it can generate a program, typically in Python, run it in a sandbox, observe the result, and iterate until the task is done; a model that works in this second way is called a \textit{code-interpreter agent} \cite{hong2024datainterpreter}. Frontier assistants offer both modes at once, and which one is used for a given dataset is decided by the model itself.

Convenient as this is for the user, LLM-based data analysis suffers from several drawbacks and limitations:
\begin{enumerate}
    \item\label{lim:cost} \textit{It consumes tokens even when generating code.} Reading data into the context to analyse it by inference consumes tokens, energy, and money in proportion to the volume analysed \cite{luccioni2024power}. Generating code rather than reading the data reduces this when the data is large, but inference is still paid for, and paid again every time the analysis is run.
    \item\label{lim:nondet} \textit{Its behaviour is not deterministic.} Because the model itself decides whether to read the data directly or to generate code --- by criteria the vendors do not document \cite{openai2026dataanalysis, anthropic2026analysis} --- the same request may be handled differently from one run to the next, and may yield different answers.
    \item\label{lim:exposure} \textit{The data leaves the organisation and reaches the model.} Files are uploaded to the provider's execution environment, and the outputs of each execution (previews of the data, printed values, error messages, intermediate results) are fed back into the model's context so that it can decide what to do next. Even when the model does not read the whole dataset, it routinely sees parts of it, and nothing in the architecture bounds what it may see. This is problematic for confidential, personal, or regulated data \cite{fernholz2025ethics}.
    \item\label{lim:disposable} \textit{The generated code is disposable.} It is a by-product of a conversation, not an artifact meant to be kept: there is no durable, inspectable program that one could store, audit, and re-run on new data of the same structure. Analysing next month's file, a new statement, or a new area means starting another conversation from scratch. This is not marginal: an accounting of one developer's coding-agent sessions over eight months found that about a quarter of all the code the agents produced --- some 350{,}000 lines --- was \textit{ephemeral}, written once, run once, and never saved \cite{monperrus2026ephemeral}.
    \item\label{lim:colocation} \textit{Model, execution, and data are colocated at the provider.} The architecture assumes that the model, the sandbox, and the data live in the same cloud environment, which makes it difficult to keep the data on premises, to call internal systems, or to use a model the organisation hosts itself.
\end{enumerate}

Our own experience with LLM-based data analysis, and discussions with developers who had built similar solutions in other contexts, led us to a common way of overcoming these issues. These solutions were each motivated by one or another of the problems above --- for example, to curb the cost of sending too much data to the LLM, or to keep sensitive data from leaving the organisation. Their root idea is a principle that privacy-sensitive data analysis adopted long before LLMs: \textit{bring the code to the data, not the data to the analyst} \cite{gaye2014datashield, williamson2020opensafely}. Because the same idea recurred across contexts, we came to define an architectural pattern that applies this principle to LLM-assisted analysis, with the LLM in the role of the analyst who never sees the data. 
In a nutshell, the LLM is given only a description of the data --- its \textit{metadata} --- and never the data itself; we call this separation the \textit{inference boundary}. The LLM answers the user's request with a program, which we call a \textit{formula}, and this formula is run where the data is. The formula is then kept, so that the same analysis can be repeated on new data without asking the LLM again. Section~\ref{sec:example} walks through these steps on a small public dataset before Section~\ref{sec:pattern} describes the pattern in full.

By instantiating this pattern, a programmer may remove or mitigate each of the limitations above. Since only metadata reaches the model, the exposure of the data is bounded (limitation~\ref{lim:exposure}). Since the formula is stored and re-run rather than regenerated, repeated analyses need no further inference, give the same results, and rely on a program that can be read and audited (limitations~\ref{lim:cost}, \ref{lim:nondet} and~\ref{lim:disposable}). And since writing the formula is separate from running it, the model can be anywhere, including on the user's own machine, while the data stays where it is (limitation~\ref{lim:colocation}).

We call the resulting analysis \textit{sustainable} in the broad sense that software engineering gives to the term, which covers environmental, economic, social, and technical dimensions \cite{lago2015framing}. Inference, and hence energy and cost, is spent once per analysis rather than once per run, in line with the call for a more sustainable use of AI itself \cite{vanwynsberghe2021sustainable}; the data stays under the control of its owner; and analyses become durable, reproducible programs rather than disposable code.

The pattern generalises text-to-SQL, in which a model writes a query from a database schema alone \cite{yu2018spider}, to arbitrary programs over tables, documents, and APIs --- as discussed in Section~\ref{sec:background} along with other related work. As such an idea may be instantiated in many variants, depending on the context and the constraints it must satisfy (safety, cost, explainability), we present our work as a \textit{pattern} in the sense of \cite{gamma1994design}: a recurring solution to a recurring problem, independent of any particular implementation. The pattern can be realised in many languages, runtimes, and with different LLM providers; our three case studies bear this out, each built on a different implementation of it.

The contributions of this paper are:
\begin{enumerate}
    \item A description of Metadata-Driven AI as an architectural pattern: context, problem, forces, participants, collaborations, variants, and consequences (Section~\ref{sec:pattern}).
    \item A positioning of the pattern with respect to code-interpreter agents (Section~\ref{sec:vs-ci}), and a cost model 
comparing the pattern with current practice and showing that inference becomes independent not only of the amount of data but also of the number of runs, together with a measurement of the information that crosses the inference boundary on public datasets (Section~\ref{sec:cost}).
    \item Three case studies that apply the pattern to different kinds of data: live geospatial APIs (Section~\ref{sec:geo}), semi-structured documents, PDF statements and spreadsheet exports (Section~\ref{sec:pdf}), and dashboard data in a model-driven smart-city application (Section~\ref{sec:city}).
    \item A discussion of the limitations of the pattern, including the information that metadata still carries, the evolution of the data and of external APIs, and the governance of reusable formulas (Section~\ref{sec:discussion}).
\end{enumerate}

\section{Metadata-Driven AI by Example}
\label{sec:example}

Before describing the pattern in general terms, we walk through one small analysis to show what it means in practice. The data is \texttt{birdstrikes}, a public table from the vega-datasets collection \cite{vegadatasets}: 10,000 wildlife-strike reports filed by US airports between 1990 and 2002, with 14 columns (airport, aircraft, date, species, phase of flight, repair cost, and so on), 1.2~MB as CSV. The user uploads the file into the client application and types the intent:

\begin{quote}
\textit{Which five wildlife species caused the highest total damage cost, and how many strikes does each account for?}
\end{quote}

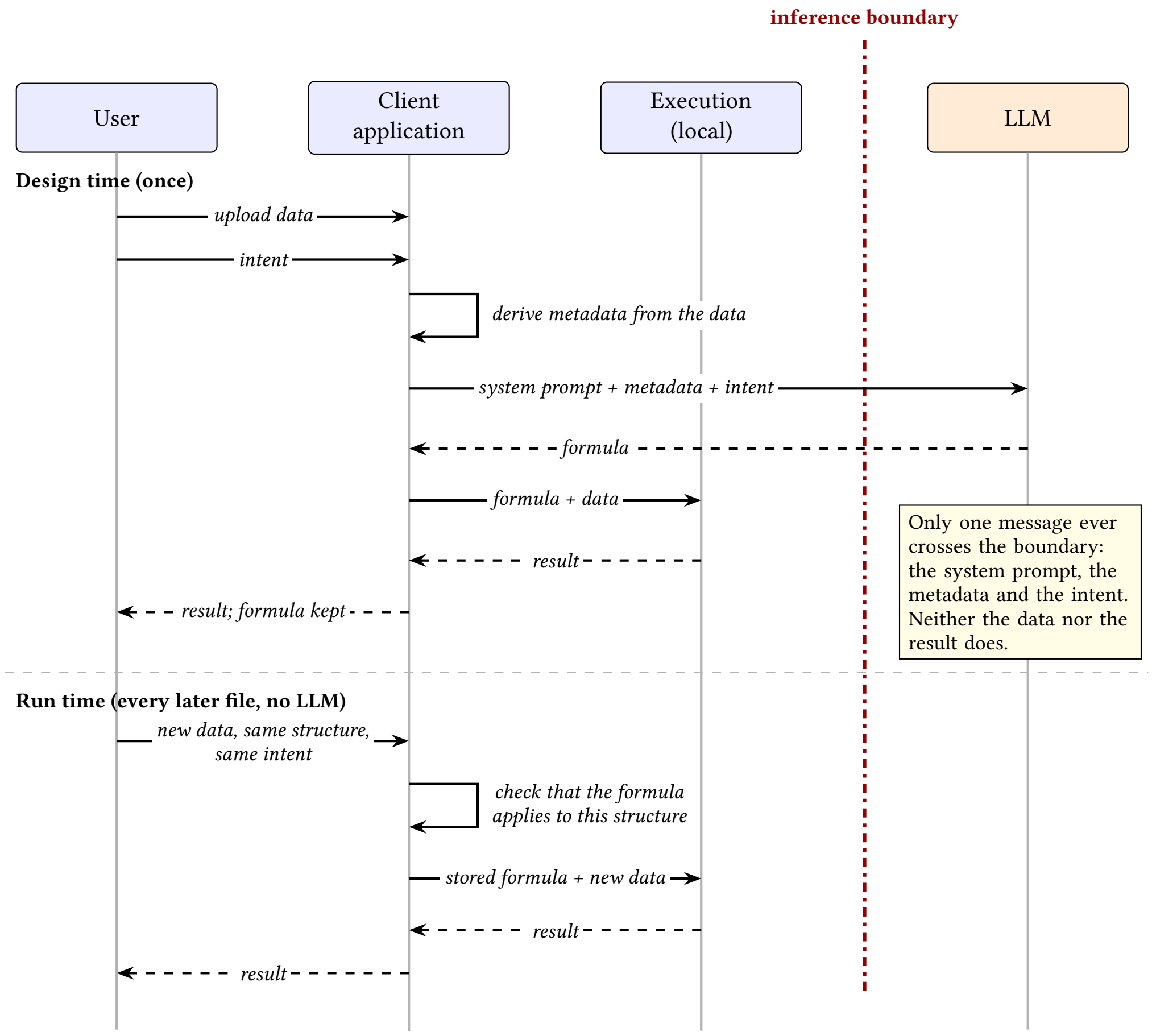
\begin{figure}
    \centering
\begin{tikzpicture}[x=1cm, y=1cm,
  head/.style={draw, rounded corners=2pt, align=center, minimum height=0.8cm, text width=2.1cm, font=\footnotesize, fill=blue!9},
  exthead/.style={head, fill=orange!16},
  msg/.style={-{Stealth[length=2mm]}, thick},
  rmsg/.style={-{Stealth[length=2mm]}, thick, dashed},
  mlbl/.style={font=\scriptsize\itshape, align=center, inner sep=1.5pt, fill=white},
  note/.style={font=\scriptsize, align=left, draw, fill=yellow!12, inner sep=3pt, text width=2.6cm},
  act/.style={fill=blue!20, draw=blue!50!black, line width=0.3pt}]
  \node[head]    (user) at (0, 0)    {User};
  \node[head]    (cli)  at (3.4, 0)  {Client\\application};
  \node[head]    (exec) at (6.8, 0)  {Execution\\(local)};
  \node[exthead] (llm)  at (10.6, 0) {LLM};
  \foreach \n in {user, cli, exec, llm} \draw[gray!60, thick] (\n.south) -- ++(0, -10.2);
  \draw[red!65!black, very thick, dash dot] (8.7, 0.9) -- (8.7, -10.2);
  \node[font=\scriptsize\bfseries, red!65!black] at (8.7, 1.15) {inference boundary};
  \node[font=\scriptsize\bfseries, anchor=west] at (-1.3, -0.75) {Design time (once)};
  \draw[msg] (0, -1.15) -- node[mlbl] {upload data} (3.4, -1.15);
  \draw[msg] (0, -1.65) -- node[mlbl] {intent} (3.4, -1.65);
  \draw[msg] (3.4, -2.05) -- ++(0.8, 0) -- ++(0, -0.5) -- (3.4, -2.55);
  \node[mlbl, anchor=west] at (4.3, -2.3) {derive metadata from the data};
  \draw[msg] (3.4, -3.15) -- node[mlbl, pos=0.35] {system prompt + metadata + intent} (10.6, -3.15);
  \draw[rmsg] (10.6, -3.85) -- node[mlbl, pos=0.7] {formula} (3.4, -3.85);
  \draw[msg] (3.4, -4.45) -- node[mlbl] {formula + data} (6.8, -4.45);
  \draw[rmsg] (6.8, -5.15) -- node[mlbl] {result} (3.4, -5.15);
  \draw[rmsg] (3.4, -5.75) -- node[mlbl] {result; formula kept} (0, -5.75);
  \draw[gray!60, dashed] (-1.3, -6.45) -- (12.0, -6.45);
  \node[font=\scriptsize\bfseries, anchor=west] at (-1.3, -6.8) {Run time (every later file, no LLM)};
  \draw[msg] (0, -7.25) -- node[mlbl] {new data, same structure,\\same intent} (3.4, -7.25);
  \draw[msg] (3.4, -7.75) -- ++(0.8, 0) -- ++(0, -0.5) -- (3.4, -8.25);
  \node[mlbl, anchor=west] at (4.3, -8.0) {check that the formula\\applies to this structure};
  \draw[msg] (3.4, -8.85) -- node[mlbl] {stored formula + new data} (6.8, -8.85);
  \draw[rmsg] (6.8, -9.45) -- node[mlbl] {result} (3.4, -9.45);
  \draw[rmsg] (3.4, -9.95) -- node[mlbl] {result} (0, -9.95);
  \node[note, anchor=north west] at (9.1, -4.5) {Only one message ever crosses the boundary: the system prompt, the metadata and the intent. Neither the data nor the result does.};
\end{tikzpicture}
    \caption{A typical Metadata-Driven analysis sequence. In the bird-strike case, the client application derives at design time 4~KB of metadata from the 1.2~MB table the user uploaded and sends it to the LLM together with a system prompt and the user's intent; the LLM returns a formula, which is executed on the real data, on the user's machine, and kept. At run time, a new file with the same structure is processed with the stored formula for the same intent, without any call to the LLM.}
    \label{fig:nutshell}
\end{figure}

Figure~\ref{fig:nutshell} shows what happens next, and what crosses the inference boundary.

\paragraph{Step 1: metadata, not data} The client derives a description of the table that is sufficient to write code against it, and sends \textit{that} to the LLM. For each column it gives the type and, for numbers, the range; for low-cardinality columns, the list of values; for other strings, the number of distinct values and three examples. Listing~\ref{lst:nutshell-meta} shows an excerpt. The whole description is 4,018 bytes, about 1,000 tokens, three hundred times smaller than the table. It is not free of information (the species names and the cost range are real values), but no row of the table is in it.

\begin{lstlisting}[numbers=none, label={lst:nutshell-meta}, caption={Excerpt of the metadata derived from the \texttt{birdstrikes} table (abridged).}]
### "birdstrikes.csv"
Rows: 10000
- Airport Name: categorical - ["ATLANTA INTL", "ATLANTIC CITY INTL", ...]
- Flight Date: string - 3625 distinct values (e.g. "1990-01-08", ...)
- Origin State: categorical - ["Arizona", "California", ..., "Washington"]
- Phase of flight: categorical - ["Approach", "Climb", "Descent", ...]
- Wildlife Species: categorical - ["American crow", ..., "Canada goose", ...]
- Cost Repair: number - range 0 ... 7043545
- Cost Total $: number - range 0 ... 7043545
- Speed IAS in knots: number - range 0 ... 350, 2836 null/empty
\end{lstlisting}

\paragraph{Step 2: the LLM writes a formula} The LLM receives a system prompt that describes the environment the program will run in (here: JavaScript, with the table available as an array of rows and a few helpers such as \texttt{parseNum}), followed by the metadata and the intent. It is asked to write a program, not an answer, and it cannot do otherwise: it has no data to answer from. Listing~\ref{lst:nutshell-formula} shows the formula it returns. The formula names the two columns it depends on; this declaration is what later tells whether it can be applied to another file.

\begin{lstlisting}[language=JavaScript, label={lst:nutshell-formula}, caption={The formula generated for the intent above.}]
// Depends on columns: "Wildlife Species", "Cost Total $"
const bySpecies = {};
for (const row of data) {
  const s = row["Wildlife Species"];
  const c = parseNum(row["Cost Total $"]) || 0;
  bySpecies[s] = bySpecies[s] || { species: s, strikes: 0, cost: 0 };
  bySpecies[s].strikes += 1;
  bySpecies[s].cost += c;
}
return Object.values(bySpecies)
  .sort((a, b) => b.cost - a.cost)
  .slice(0, 5);
\end{lstlisting}

\paragraph{Step 3: the formula runs on the real data} 
The program is executed next to the data, on the user's machine, over all 10,000 rows. Nothing has to be installed or configured for this: in our implementation, the client application runs the formula itself, in an isolated part of the browser page. It returns the answer: Canada geese account for the highest total cost, \$15.4~million over 190 strikes, followed by unidentified large birds (\$10.4~million, 436 strikes), unidentified medium birds (\$4.2~million but 3,372 strikes), rock pigeons (\$3.3~million, 232) and red-tailed hawks (\$2.3~million, 100). The result is shown to the user and is not sent back to the LLM. If the user wants a chart instead of a table, or the same ranking per phase of flight, they refine the intent and a new formula is generated, again from the metadata alone.

\paragraph{Step 4: the formula is kept} The formula is a few lines of readable code, so the user can check what it computes (for instance, that empty costs count as zero). Once validated, it is stored with the intent that produced it. When the next year's reports arrive in a file with the same structure, the stored formula is applied to them directly: the client application checks that the two columns the formula depends on are present, runs it, and shows the new ranking, without any inference. If a column were renamed, the check would fail before any wrong result is produced, and a short design-time session would repair the formula. Subtler changes, such as costs now reported in thousands of dollars, would not be caught by this check; we return to this limitation in Section~\ref{sec:evolution}. Inference was paid once, for the generation; every later run is free, deterministic, and keeps the data where it is.

\section{Background and Related Work}
\label{sec:background}

As our pattern proposal is a generalisation of a design solving recurrent problems, it is not surprising to meet related work across the literature. We try to give an overview across time and contexts of similar ideas or people working on similar problems.

\subsection{Bringing the Code to the Data}

When data cannot be moved or shown, analysis can still be performed by sending the analysis to the data. DataSHIELD \cite{gaye2014datashield} implements this for federated epidemiological studies: analysts issue commands that run where each cohort's data is held and receive only non-disclosive aggregate results. OpenSAFELY \cite{williamson2020opensafely}, which was used to analyse the primary-care records of millions of patients in England during the COVID-19 pandemic, takes the same approach with general-purpose code. Researchers develop their analysis against randomly generated dummy data that has the structure of the real records, submit the code, which is executed in the secure environment holding the records, and receive only outputs that have been checked to contain no identifiable information. The code itself is published, which makes analyses reproducible and auditable.

These systems embody the data-minimisation and privacy-by-design principles of the GDPR \cite{gdpr2016}: the analyst receives only what is necessary for the task. Metadata-Driven AI transposes their structure to LLM-assisted analysis. The metadata given to the Generator plays the role of the dummy data and data documentation given to the researcher; the Executor plays the role of the secure environment; the Orchestrator's control over which execution results, if any, are returned to the model, and in what form, plays the role of output checking; and the Formula Repository plays the role of the published code. What changes is who writes the code: a model rather than a trained researcher, which makes the approach available to users who cannot program.

\subsection{Code-Generating AI for Data Analysis}

The use of LLMs in data systems has evolved from classification tasks to agentic workflows that plan, call tools, and write code \cite{li2023autonomous}. LLMs have shown strong abilities to generate executable code \cite{austin2021program}, including for geospatial platforms such as QGIS and Google Earth Engine \cite{mai2023opportunities}, and delegating computation to generated programs makes their answers more accurate than reasoning in natural language \cite{gao2023pal}.

Text-to-SQL is the oldest and best-studied instance of generating analysis code from metadata: the model receives the schema of a database and a question, and produces a query that the database executes \cite{yu2018spider}. Because the schema is all the model needs, text-to-SQL already keeps most of the data out of the model's context; recent work further masks sensitive values that may appear in the prompt \cite{abedini2025masksql}. \textit{Ask Safely} carries the same idea to knowledge graphs: it detects sensitive values with a trained recogniser, masks them in the user's question, has the LLM generate a Cypher query from the graph schema alone, and substitutes the real values back into the query afterwards \cite{dalle2025ask}. These systems share our principle of sending the model a description of the data rather than the data itself. Our Metadata-Driven AI pattern extends it beyond SQL and any specific context: it generalises to arbitrary programs that combine local data, documents, and external APIs. Additionally, these systems are invoked when the user explicitly asks for a query, while our pattern enforces the inference boundary systematically, for every intent, whether or not the user thinks in terms of code. 

Code-interpreter agents such as Data Interpreter \cite{hong2024datainterpreter}, and the code-execution features of commercial assistants, generalise this approach to data science: the model plans, writes code, executes it in a sandbox, inspects the output, and repairs or extends the code. This loop is what makes these agents effective, but it relies on returning execution outputs, and therefore fragments of the data, to the model. In these systems, the generated code is also an ephemeral by-product of a conversation. Users must prompt the model again to repeat an analysis on new data, which hinders reproducibility and makes cost proportional to usage.

The reuse of generated code has been studied in the \textit{LLMs as tool makers} framework, where a capable model writes a tool once, as a function, and a cheaper model then applies it to many instances of the task, which amortises the cost of the capable model \cite{cai2024toolmakers}. The Metadata-Driven AI allows the user to push this amortisation further: once validated, a formula can be saved and applied again without any model.

As a final related line of work, we can mention \textit{programming by example}, where a program is synthesised from a few input--output examples and then applied to the remaining data, as popularised by FlashFill for spreadsheet string transformations \cite{gulwani2011automating}. Metadata-Driven AI shares the idea that a small amount of evidence is enough to produce a program that generalises to the rest of the data, but it replaces the domain-specific synthesis algorithm with a general-purpose LLM and uses natural language to state the intent.

\subsection{The Computational and Privacy Costs of Continuous Inference}

Continuous, cloud-based LLM inference has a significant computational and environmental cost: generating text with large models consumes orders of magnitude more energy than running conventional deterministic code \cite{luccioni2024power}. The \textit{Green AI} and \textit{Frugal AI} literature calls for efficiency to become a first-class evaluation criterion and for heavy computation to be used only where it is needed \cite{schwartz2020green, bolon2024review}. Cloud-centric workflows also raise data sovereignty issues, since analyses often involve commercial, personal, or otherwise sensitive data that data-governance policies forbid sending to third parties \cite{fernholz2025ethics}.

Local-First software prioritises the user's device as the primary location of data and computation \cite{kleppmann2019local}. Empirical studies comparing local and cloud execution show that running workloads locally can reduce the overall energy footprint by avoiding network transfers and cloud overheads \cite{siffre2025local}, and Local-First design has been framed explicitly as a Green IT strategy \cite{siffre2025local2}. Metadata-Driven AI applies this reasoning to AI-assisted analysis: the costly reasoning step is performed once, on metadata, and the repeated work runs locally. The same economics are reported for enterprise workflows by Compiled AI \cite{trooskens2026compiled}: once code is generated and validated, each execution costs no inference tokens, breaking even with a runtime-inference agent after roughly seventeen transactions and cutting token consumption 57-fold at a thousand.

\subsection{Explainability and Determinism}

Neural models are often criticised as black boxes whose reasoning cannot be traced \cite{pradhan2023spatial, jena2023explainable}, which slows their adoption in regulated domains \cite{xing2023challenges, das2020opportunities}. When the LLM produces source code instead of results, the code itself becomes the explanation: an analyst can read it, test it, and correct it \cite{austin2021program}. Once validated, its execution is deterministic and free of hallucinations; any error is a programming error that can be found and fixed.

Compiled AI \cite{trooskens2026compiled} builds a whole enterprise-workflow system around this determinism: an LLM generates business-logic code once, during a ``compilation'' phase, and every artifact must pass a mandatory multi-stage validation pipeline --- security analysis, syntactic checks, execution testing, and output-accuracy checks --- before it is deployed and then run with no further model calls. It is the closest work to our pattern, arrived at independently in the context of high-stakes workflow automation (and evaluated on function-calling and invoice processing for healthcare and CRM settings). The two designs share the same backbone --- generate code once, validate it, and re-execute it deterministically without inference --- but differ in what organises the architecture: Compiled AI is organised around determinism and auditability, whereas Metadata-Driven AI is organised around the one-way inference boundary, so that it is the data itself, and not merely the cost or variability of inference, that is kept away from the model. In this sense Compiled AI can be read as an instantiation of the same idea specialised to enterprise workflows.

\section{The Metadata-Driven AI Pattern}
\label{sec:pattern}

In this section, we describe our Metadata-Driven AI pattern using the usual structure of design-pattern catalogues \cite{gamma1994design}.

\subsection{Intent, Context, and Problem}

\paragraph{Intent} Let users analyse data in natural language with the help of an LLM, without sending the data to the LLM, and turn each analysis into a deterministic program that can be reused on new data without further inference.

\paragraph{Context} An organisation holds data that it cannot or does not want to send to an LLM provider --- customer records, financial documents, health or city data --- or that is too large to fit in a model's context (geospatial layers, transaction histories, and clinical records are all routinely this large). Even when the data does fit, sending it on every analysis costs money and, if every organisation does the same, carries an avoidable environmental cost. Users who are not programmers want to ask questions about it, produce charts, extract fields, or enrich it with external data. The data is not necessarily tabular: a set of documents that share a layout, such as the monthly statements of one bank or the invoices of one supplier, is data of the same kind for the pattern, with the structure of one sample playing the role of the schema, even when the documents themselves contain tables to be extracted. The same analyses tend to be repeated on new data of the same shape: new rows, new documents, new time periods, or new slices of a domain --- for instance new geographic areas in a GIS setting, or new accounts or reporting periods elsewhere.

\paragraph{Problem} How can the reasoning abilities of an LLM be used for such analyses while keeping the data out of the model's context, making results reproducible across runs, and paying for inference once per analysis rather than once per run?

\paragraph{Forces} These analyses are shaped by competing concerns that any solution must balance:
\begin{itemize}
    \item \textit{Expressiveness vs.\ exposure}: the model needs enough of a description to write correct code, and richer metadata can sometimes help, but every added detail exposes more of the data.
    \item \textit{Feedback vs.\ exposure}: a model that observes the results of its code can correct itself, but these results are derived from the data.
    \item \textit{Flexibility vs.\ determinism}: natural language is flexible, but analytical results must be repeatable and auditable.
    \item \textit{Freshness vs.\ replication}: joining local data with live external sources is valuable, but replicating these sources locally (ETL) is costly to build and maintain.
    \item \textit{Capability vs.\ sovereignty}: the largest models are cloud-hosted, while sovereignty favours models that can run on premises or on the user's device.
\end{itemize}

\subsection{Solution}

Separate the analysis into a \textit{design time}, where the LLM generates a program from metadata, and a \textit{run time}, where the program is executed locally on the real data without the LLM. The boundary between what the model sees and what it does not see (the \textit{inference boundary}) is enforced by construction: the participant that talks to the LLM only ever receives metadata.

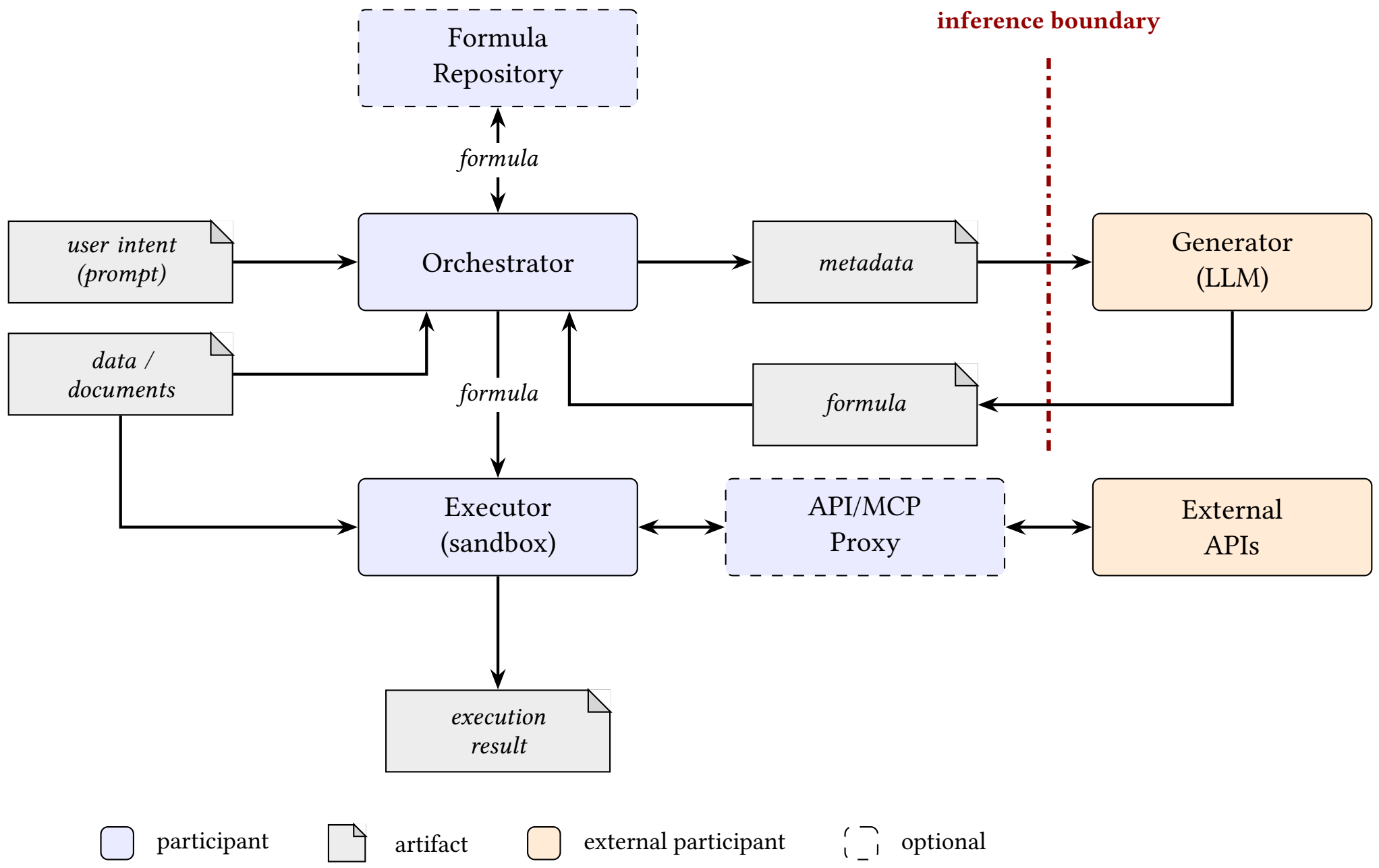
\begin{figure}
    \centering
\begin{tikzpicture}[x=1cm, y=1cm,
  comp/.style={draw, rounded corners=2pt, align=center, minimum height=0.95cm,
               text width=2.5cm, font=\footnotesize, fill=blue!9},
  art/.style={draw, fill=gray!14, align=center, font=\scriptsize\itshape,
              inner sep=3.5pt, minimum height=0.8cm, text width=1.95cm},
  ext/.style={comp, fill=orange!16},
  opt/.style={dashed},
  arr/.style={-{Stealth[length=2mm]}, thick},
  biarr/.style={{Stealth[length=2mm]}-{Stealth[length=2mm]}, thick},
  lbl/.style={font=\scriptsize\itshape, align=center, inner sep=1.5pt, fill=white}]

  \node[comp, opt] (repo) at (0, 3.0)  {Formula\\Repository};
  \node[art]  (user)  at (-3.7, 1.0)  {user intent\\(prompt)};
  \node[comp] (orch)  at (0, 1.0)     {Orchestrator};
  \node[art]  (data)  at (-3.7, -0.1) {data /\\documents};

  \node[art]  (meta)  at (3.6, 1.0)  {metadata};
  \node[art]  (form)  at (3.6, -0.4) {formula};
  \node[ext]  (gen)   at (7.2, 1.0)  {Generator\\(LLM)};

  \node[comp] (exec)  at (0, -1.6)   {Executor\\(sandbox)};
  \node[art]  (res)   at (0, -3.6)   {execution\\result};
  \node[comp, opt] (proxy) at (3.6, -1.6) {API/MCP\\Proxy};
  \node[ext]  (apis)  at (7.2, -1.6) {External\\APIs};

  \fold{user}\fold{data}\fold{meta}\fold{form}\fold{res}

  \draw[arr] (user.east) -- (orch.west);
  \draw[biarr] (repo.south) -- node[lbl] {formula} (orch.north);
  \draw[arr] (orch.east)  -- (meta.west);
  \draw[arr] (meta.east)  -- (gen.west);
  \draw[arr] (gen.south)  |- (form.east);
  \draw[arr] (form.west)  -| ([xshift=0.7cm]orch.south);

  \draw[arr] (data.east) -| ([xshift=-0.7cm]orch.south);

  \draw[arr] (orch.south) -- node[lbl] {formula} (exec.north);
  \draw[arr] (data.south) |- (exec.west);
  \draw[arr] (exec.south) -- (res.north);
  \draw[biarr] (exec.east)  -- (proxy.west);
  \draw[biarr] (proxy.east) -- (apis.west);

  \draw[red!65!black, very thick, dash dot] (5.4, 3.0) -- (5.4, -0.9);
  \node[font=\scriptsize\bfseries, red!65!black, align=center] at (5.4, 3.35) {inference boundary};

  \node[comp, minimum height=0.32cm, text width=0.32cm, inner sep=0pt, anchor=west] (k1) at (-3.9, -4.7) {};
  \node[font=\scriptsize, right=0.1cm of k1] (k1t) {participant};
  \node[draw, fill=gray!14, minimum height=0.36cm, text width=0.36cm, inner sep=0pt, right=0.45cm of k1t] (k2) {};
  \fold[0.12]{k2}
  \node[font=\scriptsize, right=0.16cm of k2] (k2t) {artifact};
  \node[ext, minimum height=0.32cm, text width=0.32cm, inner sep=0pt, right=0.45cm of k2t] (k3) {};
  \node[font=\scriptsize, right=0.1cm of k3] (k3t) {external participant};
  \node[draw, dashed, rounded corners=2pt, minimum height=0.32cm, text width=0.32cm, inner sep=0pt, fill=white, right=0.45cm of k3t] (k4) {};
  \node[font=\scriptsize, right=0.1cm of k4] (k4t) {optional};
\end{tikzpicture}
    \caption{Participants and artifacts of the Metadata-Driven AI pattern. }
    \label{fig:pattern}
\end{figure}

\paragraph{Participants and artifacts} Figure~\ref{fig:pattern} shows the structure of the pattern. As in pattern catalogues \cite{gamma1994design}, we describe it through its \textit{participants} and their responsibilities. As in \cite{perry1992foundations, garlan1993introduction}, we keep the participants separate from the \textit{artifacts} they produce and consume, that is, the data. A participant is a responsibility, not necessarily a software module, and some participants are optional for the pattern to work.

\textit{Participants.}
\begin{itemize}
    \item \textbf{Orchestrator}: the core of the pattern and the only participant that crosses the inference boundary to the Generator. It derives the metadata from the data sources and composes the generation request --- a system prompt that describes the execution environment, the metadata, the user's intent, and optionally examples retrieved from the Formula Repository. It checks the generated formula and drives the two feedback loops: an (optional) automatic \textit{retry}, in which a formula that fails is regenerated without the user's involvement, and a \textit{refinement}, in which the logs and errors of the previous execution accompany the user's next intent. It also enforces the inference boundary and makes it auditable: it logs what crosses, and controls what is sent back to the Generator --- metadata by default, and execution feedback (error messages, logs, a sample of the output) only when the application allows it. Last but not least, it may implement some obfuscation mechanism on the intent, the metadata, and the logs to avoid leaking confidential data (as done, for instance, for query generation \cite{dalle2025ask, abedini2025masksql}). Since it has access to the data, it can also check the intent itself: if the user's prompt contains values copied from the data, for instance a pasted excerpt, it can refuse to send it or ask the user to rephrase. It is up to the designer to decide whether these decisions are built in, configuration-driven, or user-driven, depending on the context. Note that the Orchestrator may run anywhere trusted by the data owner, on the user's device or on an organisational gateway; the pattern does not prescribe where.
    \item \textbf{Generator}: the LLM. Given the request composed by the Orchestrator, its sole responsibility is to return a formula and an explanation of what the formula does. It never sees the data, only metadata, and it can run anywhere: a frontier cloud model, a model on the organisation's premises, or a small model on the user's device (Section~\ref{sec:smallmodels}), which is what makes fully sovereign deployments possible.
    \item \textbf{Executor}: runs formulas on the real data and returns the result. When formulas are code, it is a sandbox; when they are declarative (see \textit{Variants}), it is a fixed, trusted function that interprets them, and no sandbox is needed. In our implementation, it is a sandboxed browser \texttt{iframe} that runs on the user's device, has no direct network access, and receives the data from the host application; external calls, when a formula makes any, go through the API Proxy. The Executor also provides the libraries and helper functions that a formula may use (charting, geometry, document and number parsing, and others), whose availability is described to the Generator in the system prompt.
    \item \textbf{API Proxy} (optional): the only outbound channel of the Executor, needed when formulas call external services. It restricts calls to an allow-list, holds the service credentials so that they never appear in a formula, and logs what leaves. A Model Context Protocol \cite{anthropic2024mcp} client is a natural instantiation of this role. The responsibility can be fulfilled by a proxy process (necessary for a browser sandbox, which has no network access of its own) or by the egress policy and secret store of a server-side Executor.
    \item \textbf{Formula Repository} (optional): stores validated formulas with the intent that produced them, their title, description, and declared dependencies. It is what makes run-time reuse possible and what supplies the Orchestrator with examples relevant to a new intent --- a form of retrieval-augmented generation \cite{lewis2020retrieval}, close to the selection of in-context examples by semantic similarity \cite{liu2022makes}. An instantiation in which formulas are only used interactively, at design time, can omit it.
\end{itemize}

\textit{Artifacts.}
\begin{itemize}
    \item \textbf{Data and documents}: the actual analysis target --- local tables, documents, and databases. They are read only by the Orchestrator, to derive the metadata, and by the Executor, to run the formula on; they never cross the inference boundary. They need not be local: they may reside wherever the Orchestrator does, as long as that is trusted by the data owner. For documents, the unit of data is the document, and a family of documents with a common structure (the statements of one bank, for instance) is treated like a table whose rows are the documents: the formula is written once for the structure and applied to every document that has it.
    \item \textbf{User intent}: the natural-language prompt that states what the user wants to obtain from the data (for example: \emph{What is the average of a quantity over the last year?}). Note that the intent may be obfuscated to avoid leaking confidential data \cite{dalle2025ask}.
    \item \textbf{Metadata}: the description of the data that is sufficient to write code against it --- column names and types, value ranges, category lists, the number of rows; for documents, a structure-preserving rendering of a sample; for APIs, their contracts (in the bird-strike example of Section~\ref{sec:example}, it is the 4~KB listing of column types, ranges and category lists). It is the only representation of the data that crosses the inference boundary, but it can be obfuscated by the Orchestrator \cite{abedini2025masksql}.
    \item \textbf{Formula}: the program that fulfils the user's intent over the data. It is deterministic and human-readable. In the bird-strike example it is a short JavaScript program (Listing~\ref{lst:nutshell-formula}), but the pattern does not prescribe its form: it may be a program in any language, a query such as SQL, or a declarative specification that a trusted function of the Executor interprets instead of executing code (see \textit{Variants} below); what matters is that the Executor can run it. It declares the structure it depends on (the columns, datasets or document anchors it relies on), so that its applicability to new data can be checked without running it. By default it returns the structured answer to the intent; when the context calls for a rendering, such as a chart or a map, the system prompt asks it to produce one with the Executor's libraries. It is the pattern's first-class, persistent artifact: generated once, checked by the Orchestrator, executed on the real data, and stored for reuse without further inference.
    \item \textbf{Execution result}: what the formula returns when run on the real data --- the structured answer to the intent. It is produced entirely by the Executor and, by default, is not shown to the Generator.
\end{itemize}

\paragraph{Who plays the Orchestrator?} The Orchestrator is a responsibility, and who fulfils it shapes both the guarantees of the pattern and what it asks of the user. In the implementations we describe, it is deterministic software --- the client application of Section~\ref{sec:example} --- and the metadata is derived automatically by an extractor written for the kind of data at hand; the user only states the intent and judges the result. This keeps the boundary strict, since no model sees the data, but each new kind of data needs its own extractor, which is easy for tables and harder for semi-structured documents. Part of the role can also be played by a human: an analyst who writes or curates the metadata by hand, for instance to describe data that no extractor supports, or to leave out sensitive values. This requires abstraction skills --- describing data well enough for code to be written against it, without revealing it --- that a user who simply hands the data to an LLM does not need. Finally, the Orchestrator may itself be an agent, which is attractive for deriving metadata from arbitrary data or for deciding which feedback to return. But an agent that derives metadata has to read the data, which breaks the boundary if that agent is the Generator or runs outside the trusted space. An agentic Orchestrator is therefore compatible with the pattern only if its model runs inside the trusted space, for instance a small local model. Two models then collaborate: one that sees the data and stays with it, and a Generator, possibly larger and remote, that only sees what the first one lets through.

\begin{figure}
    \centering
\begin{tikzpicture}[node distance=0.45cm]
  \node[box] (p) {State intent};
  \node[local, below=of p] (m) {Extract metadata};
  \node[llm, below=of m] (g) {Generate formula};
  \node[local, below=of g] (e) {Execute on real data};
  \node[box, below=of e] (v) {Inspect result};
  \node[local, below=of v] (s) {Save formula};
  \draw[arr] (p) -- (m); \draw[arr] (m) -- (g); \draw[arr] (g) -- (e); \draw[arr] (e) -- (v);
  \draw[arr] (v) -- node[lbl, right] {validated} (s);
  \draw[arr] (v.west) -- ++(-0.5,0) |- node[lbl, pos=0.25, left] {refine\\or fix} (p.west);
  \draw[darr] (e.east) -- ++(0.5,0) |- node[lbl, pos=0.25, right] {retry\\on error} (g.east);
  \draw[darr] ([yshift=-0.2cm]e.east) -- ++(1.7,0) |- node[lbl, pos=0.25, right] {previous\\logs and\\errors} (p.east);
  \node[lbl, font=\scriptsize, align=left, anchor=west] at ($(e.east)+(0.6,-0.85)$) {dashed: feedback\\(if allowed)};
  \node[font=\footnotesize\bfseries, above=0.15cm of p] {(a) Design time};

  \node[box, right=5.2cm of p] (q) {Select formula\\(dependency check)};
  \node[local, below=of q] (n) {Load new data};
  \node[local, below=of n] (x) {Execute on real data};
  \node[box, below=of x] (r) {Use result};
  \draw[arr] (q) -- (n); \draw[arr] (n) -- (x); \draw[arr] (x) -- (r);
  \draw[arr] (r.east) -- ++(0.5,0) |- node[lbl, pos=0.25, right] {next item,\\document,\\or period} (n.east);
  \node[font=\footnotesize\bfseries, above=0.15cm of q] {(b) Run time (no LLM)};
\end{tikzpicture}
    \caption{The two workflows of the pattern. (a) At design time, the user iterates with the Generator, which only sees metadata, until the formula produces the expected result on the real data. The dashed arrows carry execution feedback, returned only when the application allows it: in the retry loop, a failed formula is regenerated from its error and logs; in the refinement loop, the logs and errors of the previous execution accompany the user's new intent. (b) At run time, a stored formula is applied to new data of the same structure without any inference. Orange: steps that involve the LLM; green: steps that run locally.}
    \label{fig:workflows}
\end{figure}
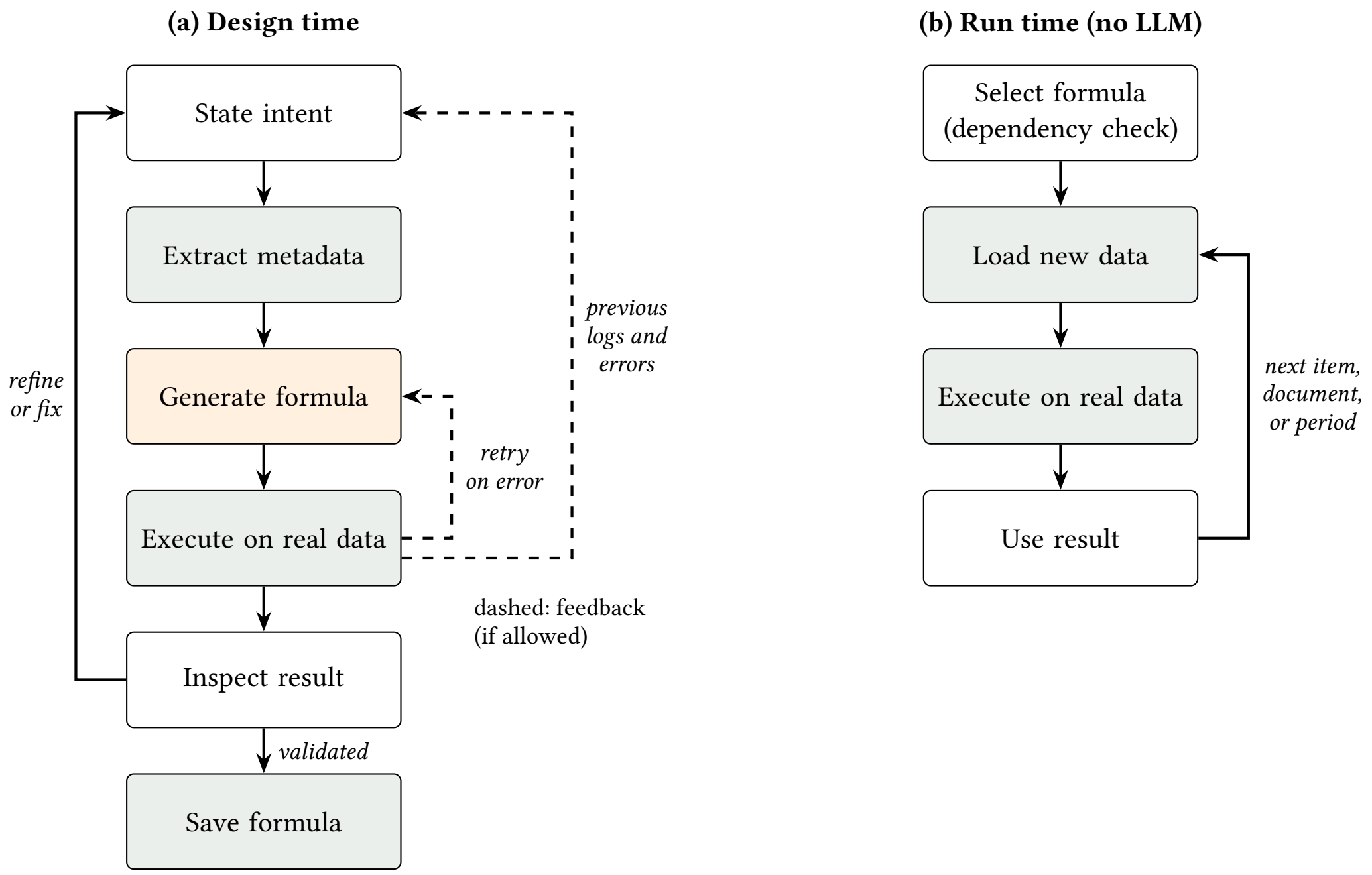

\paragraph{Collaborations} Figure~\ref{fig:workflows} shows the two workflows of the pattern. In the \textit{design-time workflow}, the user states an intent; the Orchestrator extracts the metadata and composes the request; the Generator returns a formula; the Executor runs it on the real data and shows the result; and the user refines the intent until the result is satisfactory, then saves the formula. When the application allows execution feedback, it can enter this workflow in two ways. In the \textit{retry loop}, when the execution of a formula fails, the error and the formula's logs are filtered by the Orchestrator and returned to the Generator, which produces a corrected formula; this can be repeated a bounded number of times before the result is shown to the user. In the \textit{refinement loop}, when the user is not satisfied and states a new intent, the filtered logs and errors of the previous execution are sent together with it, so that the Generator knows how the previous formula behaved. For instance, the user may say that the formula returned five items instead of six, and the logs let the Generator find which step missed the sixth one. When no feedback is allowed, the user describes the problem in their own words. In the \textit{run-time workflow}, which requires the Formula Repository, the user (or an application, or a scheduled job) selects a formula whose dependencies are satisfied by the data at hand and executes it, without any LLM call.

\paragraph{The generation request} The request the Orchestrator composes is where the inference boundary takes concrete shape: everything the Generator ever learns about the data is what this request carries. It is convenient to think of it as a system prompt with a small number of slots (Listing~\ref{lst:sysprompt}). A \textit{role} tells the Generator that its task is to write a formula rather than to answer directly, and in which form: a program in a given language, or a declarative specification conforming to a given schema. An \textit{environment} section declares the runtime, the helper libraries the formula may call (charting, geometry, document and number parsing, and others), and the contracts of the external services reachable through the API Proxy; it is what lets the Generator write code against capabilities it cannot see. The \textit{metadata} slot is the one the variants below vary: a schema with statistics, a structure-preserving exemplar, or a domain model. An optional \textit{examples} slot carries formulas retrieved from the Formula Repository for similar past intents. An \textit{output contract} instructs the Generator to return a structured answer (and a rendering when the context calls for one), to declare the structure the formula depends on, to anchor on static descriptors rather than on sample values, to check that structure first and fail fast, and not to hard-code any value taken from the metadata; it is this contract that gives the formula its reusable, auditable shape, and the case studies specialise it (the fail-fast structure check of Section~\ref{sec:pdf} is one instance). Finally, the \textit{intent} slot carries the user's request, optionally accompanied by the filtered errors and logs of the previous execution when the feedback policy allows it. The request is deliberately independent of any one implementation: the same skeleton underlies all three case studies, and only the metadata and the environment differ.

\begin{lstlisting}[numbers=none, label={lst:sysprompt}, caption={Skeleton of the generation request the Orchestrator composes. Angle brackets mark slots filled per analysis; the \textit{examples} slot and the trailing feedback are optional. Only the metadata, the API contracts, and the intent cross the inference boundary.}]
[ROLE]        Translate the user's intent into a FORMULA over the data
              described below: <a self-contained program in <language> |
              a declarative specification conforming to <schema>>.
              You never see the data itself.
[ENVIRONMENT] Runtime: <executor>. Helpers available to the formula:
              <charting, geometry, document and number parsing, ...>.
              Services reachable through the proxy: <API contracts>.
[METADATA]    <schema + statistics | structure-preserving exemplar | domain model>
[EXAMPLES]    (optional) formulas written for similar past intents:
              <intent -> formula, retrieved from the Formula Repository>
[OUTPUT]      Return a structured answer (plus a rendering when asked).
              Declare the structure you depend on. Anchor on static
              descriptors, never on sample values. Check that structure
              first and fail fast. Do not hard-code values from the metadata.
[INTENT]      <the user's natural-language request>
              (optional) errors and filtered logs of the previous execution
\end{lstlisting}

\paragraph{Variants} The pattern admits several variants, which the case studies illustrate:
\begin{itemize}
    \item \textit{Metadata granularity.} From schema-only (names and types), to schema with statistics (ranges, category lists), to a \textit{structural exemplar} (the full text of one sample document, Section~\ref{sec:pdf}), to a formal model of the domain (Section~\ref{sec:city}). Richer metadata gives the Generator more to reason about but exposes more information; the choice is a deliberate trade-off (see Section~\ref{sec:leakage}).
    \item \textit{Location of the Generator.} A cloud-hosted frontier model, a model hosted on the organisation's premises, or a small model on the user's device. Since the Generator only writes code, small models can be sufficient (Section~\ref{sec:smallmodels}), which makes fully sovereign deployments possible.
    \item \textit{Form of the formula.} In the \textit{executable} form, the Generator writes a program and the Executor is a sandbox; this is the most flexible form, since a formula can combine parsing, computation, API calls and rendering in any way, and it is the form used in the bird-strike example. In the \textit{declarative} form, the system prompt does not ask for code but for a description of the analysis in a fixed vocabulary --- a filter-group-aggregate description, a chart specification, or simply the parameters of a predefined analysis --- and the Executor is a trusted function of the application that interprets it over the data. For the bird-strike intent, instead of the program of Listing~\ref{lst:nutshell-formula}, the Generator would return something like:
\begin{lstlisting}[numbers=none]
{ "groupBy": "Wildlife Species",
  "measures": { "strikes": "count", "cost": { "sum": "Cost Total $" } },
  "orderBy": "cost", "limit": 5 }
\end{lstlisting}
which a group-and-aggregate function written once by the developers executes. The declarative form can only express what that function supports, but it is far simpler to implement, needs no sandbox, and is safe out of the box, since nothing returned by the LLM is ever executed as code; text-to-SQL, with the database engine as Executor, is its best-known instance. The rest of the pattern --- the inference boundary, the persistence of formulas and their declared dependencies, the cost model --- is the same in both forms, and an application can mix them, using declarative formulas for routine questions and executable ones where flexibility is needed.
    \item \textit{Location of the Executor.} A browser sandbox, a server-side container, an edge device, or, for declarative formulas, the application itself.
    \item \textit{Data access.} Data pushed into the Executor by the host application, or pulled just-in-time from APIs through the API Proxy.
    \item \textit{Feedback policy.} No feedback at all, the user describing problems in their own words; a retry loop limited to errors that carry no data, such as syntax errors, with a bounded number of retries; feedback of the previous execution attached to the user's next request in the refinement loop; return of the formula's logs and of a sample of its output, redacted or approved by the user; or an automatic revision step, in which the first formula generated for a new document is executed and its logs are returned once to the Generator before anything is shown to the user. In our implementation, syntax errors are returned automatically (one retry by default), whereas runtime logs and output samples are returned only if the host application explicitly passes them, and the automatic revision step for documents is disabled by default. It does not yet provide redaction.
    \item \textit{Retrieval strategy.} No retrieval at all; the nearest stored intent by semantic similarity, whose formula is given as a single example; or richer strategies, such as filtering candidates by the compatibility of their declared dependencies with the current metadata, retrieving several formulas or fragments of code (a parsing routine, an API call), or favouring formulas that have been validated and reused most often.
\end{itemize}

\subsection{Consequences}

The pattern has the following benefits:
\begin{enumerate}
    \item \textit{Data sovereignty}: the data never enters the model's context; only metadata does, plus execution feedback if the application allows it, and only in the form it permits. The Orchestrator and the API Proxy can log exactly what was sent.
    \item \textit{Reproducibility}: a validated formula always produces the same output for the same input.
    \item \textit{Explainability}: the formula is a readable artifact that can be audited, tested and corrected, rather than an opaque model output.
    \item \textit{Frugality}: inference is paid once per analysis, not once per item, per session, or per execution (Section~\ref{sec:cost}).
    \item \textit{Scalability}: execution speed is bounded by local compute, not by LLM latency or rate limits, and the size of the data is not limited by the model's context window.
    \item \textit{Adaptability}: an existing formula can be adapted to a new context by asking the Generator to modify it, which is cheaper and more reliable than generating it from scratch.
    \item \textit{Continuous improvement} (with the optional Formula Repository): validated formulas become examples for future generations, so the quality of generated formulas can improve with the quality of the stored examples, without changing the underlying model.
\end{enumerate}

\paragraph{Is the pattern still useful with a local model?} When the Generator runs on infrastructure that the data owner controls, giving it the data no longer raises a privacy issue, and one may ask whether the inference boundary is still worth enforcing. Most of the benefits above do not depend on where the model runs. Local inference is not free: it costs energy and time, and on the hardware where such models run both grow quickly with the size of the context (in our benchmark, one generation from a short request already took over 20~s on a laptop, Table~\ref{tab:bench}); the context windows of small models are also much smaller than those of frontier models, so reading a table such as \texttt{birdstrikes} into the context is often not an option. Reproducibility, explainability, and reuse without further inference come from the formula, not from the location of the model. And keeping the data out of the model's context remains a useful default within an organisation, where access to some data is restricted to some users while the model is shared. With a local model, the pattern therefore remains applicable, but the case for it rests on frugality, determinism, and reuse rather than on sovereignty.

It also has liabilities, discussed in Section~\ref{sec:discussion}: metadata is not free of information; formulas depend on external APIs that can change; client-side execution has limited resources; executable formulas need a sandbox (or a trusted interpreter); and a shared repository of formulas needs governance. Also, a stored formula can fall out of line with data whose structure or meaning changes, sometimes silently, and noticing and repairing this requires some understanding from the user. Finally, the pattern only applies to tasks that can be expressed as a program over the data. Tasks that require reading and interpreting the data itself, such as summarising free text or judging the tone of a message, need the data to be shown to a model and fall outside its scope; they can still be combined with the pattern by running a local model on the specific fields that require it.

\subsection{Comparison with Code-Interpreter Agents}
\label{sec:vs-ci}

Code-interpreter agents and Metadata-Driven AI both use the LLM to write code rather than to compute results. Table~\ref{tab:vs-ci} summarises where they differ. In short, a code-interpreter agent \textit{may} behave like the pattern (a careful agent can inspect only the schema of a file and write reusable code), but nothing in its architecture guarantees it. Metadata-Driven AI turns these behaviours into constraints that hold by construction: what the model sees and what it receives back are bounded by the Orchestrator (metadata by default, and execution feedback only when and in the form the application allows), and reuse is supported by the Formula Repository and by the formulas' declared dependencies.

\begin{table*}[t]
\centering
\caption{Code-interpreter agents vs. Metadata-Driven AI.}
\label{tab:vs-ci}
\small
\renewcommand{\arraystretch}{1.25}
\begin{tabularx}{\textwidth}{@{} >{\raggedright\arraybackslash}p{3.3cm} >{\raggedright\arraybackslash}X >{\raggedright\arraybackslash}X @{}}
\toprule
 & \textbf{Code-interpreter agent} & \textbf{Metadata-Driven AI} \\
\midrule
Where the data goes & uploaded to the provider's sandbox & stays where it is; formulas run next to it \\ \addlinespace
What the model sees & whatever the agent decides to inspect: previews, printed values, errors, intermediate results & metadata; execution results go to the user, and back to the model only when the Orchestrator allows it \\ \addlinespace
Lifecycle of the code & ephemeral, part of a conversation & persistent artifact with declared dependencies, stored and versioned \\ \addlinespace
Re-running on new data & new session, new inference, possibly different code & no inference; dependency check, then execution \\ \addlinespace
Inference grows with & number of sessions (runs) & number of distinct analyses \\ \addlinespace
Reproducibility across runs & not guaranteed & guaranteed for a given formula and input \\ \addlinespace
Access to internal systems and APIs & limited by the provider's sandbox & through the API Proxy, with an allow-list and server-side credentials \\ \addlinespace
Location of the model & provider's cloud & any (cloud, on premises, or local) \\ \addlinespace
Self-correction & automatic, by observing outputs & controlled: feedback is returned only under the Orchestrator's control (filtered or approved), or not at all \\
\bottomrule
\end{tabularx}
\end{table*}

The pattern gives up some autonomy: an agent that freely inspects outputs can explore the data and fix its own mistakes without the user's involvement. Metadata-Driven AI trades this for a bounded, auditable exposure and for reusable artifacts. The trade-off is most favourable when analyses are repeated on data of the same structure, when the data is sensitive, or when the organisation wants to choose where the model runs.

\paragraph{Why not tool use?} A natural objection is that external APIs could simply be exposed to the model as tools, through function calling or a standard such as the Model Context Protocol (MCP) \cite{anthropic2024mcp}, or packaged with instructions and scripts as a \textit{skill} \cite{anthropic2025skills}. The difference is who calls the API. With tool use, the model issues the call and the tool's result is returned into its context: the data reaches the model, and each run needs a model turn, so inference grows with the number of runs exactly as for code-interpreter agents. A skill is closer to a stored formula, since it packages reusable instructions and code, but it is executed in the provider's sandbox and its outputs are observed by the model. In the pattern, the API is called by the formula at run time, through the API Proxy, with no model involved; the model only ever sees the API's contract. Tool definitions remain useful to the pattern: the description of an MCP server's tools is a ready-made API contract, that is, metadata, and the API Proxy can act as an MCP client at run time so that formulas call the same tools without the model.

\section{Cost Model and Measurements}
\label{sec:cost}

\subsection{A Simple Cost Model}

Consider an analysis applied to $N$ items (rows, documents, geographic targets) and run $R$ times over its lifetime (for example, each month on new data, or once per new statement). The model applies equally to tables and to documents: for a family of bank statements, an item is one statement, $R$ is the number of statements processed over the life of the extractor, and the metadata is the structure-preserving text of one sample statement, larger than a table's schema but still paid once per layout rather than once per document. 
To characterise the cost advantage of the pattern, we compare its inference cost with that of the two ways in which assistants analyse data today (Section~\ref{sec:intro}): placing the data in the model's context, and code-interpreter agents. These are not variants of the pattern but the baselines against which it is measured. As noted in Section~\ref{sec:vs-ci}, a carefully used code-interpreter agent can behave like the pattern within a session; we model it as it is typically used, with code that is not kept, so that each run starts a new session. An agent whose code was stored and re-run without the model would, in effect, instantiate the pattern, and its cost would be the one we give for the pattern below.

\paragraph{Data-in-context agent} Each item is placed in the model's context, with $T_{ctx}$ tokens of instructions and tool descriptions, $T_{data,i}$ tokens of data, and $T_{out,i}$ tokens of output:
\begin{equation}
  \mathcal{T}_{dic} = R \cdot \sum_{i=1}^{N} \left( T_{ctx} + T_{data,i} + T_{out,i} \right)
\end{equation}
Cost grows with the amount of data. If we take again the bird-strike example of Section~\ref{sec:example}, the table alone is about 300,000 tokens (1.2~MB); reading it into the context exceeds the context window of most models and, even with the largest models being able to hold it, that reading would be paid again for every question and for every new file. This case is not only historical. It applies whenever the model reads the data instead of processing it with code, either because the task requires it (e.g., summarising free text) or because the assistant chooses to, for instance when a PDF statement is short enough to be read directly. Since this choice belongs to the model and is not documented (Section~\ref{sec:intro}), the user of an assistant cannot rule this case out in advance.

\paragraph{Code-interpreter agent} In each session, the agent performs $K$ generation steps. At step $k$, the context contains the instructions, a description of the data $T_{desc}$ (typically obtained by inspecting the file), the history $T_{hist,k}$, and the observations $T_{obs,k}$ returned by the previous executions; the output is code $T_{c,k}$:
\begin{equation}
  \mathcal{T}_{ci} = R \cdot \sum_{k=1}^{K} \left( T_{ctx} + T_{desc} + T_{hist,k} + T_{obs,k} + T_{c,k} \right)
\end{equation}
This cost no longer depends on $N$, except through the size of the observations, but it is paid again at each of the $R$ runs. In the bird-strike example, the agent would upload the table, inspect it (a preview of the first rows and the column list), write the same aggregation, and print its output; next year's file would mean a new session, with the same steps. Moreover, $T_{desc}$ and $T_{obs,k}$ are derived from the data: they are the channel through which data reaches the model. This formula describes the best case for the agent, in which all the data is processed by code. A session may also mix the two modes, reading some files or documents directly and processing others with code, in which case its cost lies between $\mathcal{T}_{ci}$ and $\mathcal{T}_{dic}$.

\paragraph{Metadata-Driven AI} The LLM is invoked $K$ times at design time, with the metadata $T_{meta}$ and without observations, except the execution feedback $T_{err,k}$ (errors, filtered logs) that the Orchestrator may return, which is usually short and is zero when no feedback is returned. Automatic retries are generation steps and are counted in $K$:
\begin{equation}
  \mathcal{T}_{mf} = \sum_{k=1}^{K} \left( T_{ctx} + T_{meta} + T_{hist,k} + T_{err,k} + T_{f,k} \right)
\end{equation}
In the bird-strike example, $T_{meta}$ is about 1,000 tokens (4,018 bytes) and the formula $T_{f}$ about a hundred, so with $K=1$ the entire inference cost of the analysis is one request of $T_{ctx}$ plus some 1,100 tokens, however many files the formula is later applied to. This quantity depends neither on $N$ nor on $R$. The $N \cdot R$ executions of the formula run locally, at a cost that is negligible compared to LLM inference \cite{luccioni2024power}. When examples are retrieved from the optional Formula Repository, a similar formula given as an example can reduce $K$.

For a single run ($R=1$), and when the agent processes the data with code, the code-interpreter agent and Metadata-Driven AI have comparable inference costs; the pattern's advantages are then limited to the bounded exposure of the data and the choice of where the model and the execution run. As soon as an analysis is repeated, the cost of the code-interpreter agent grows linearly with $R$, while that of Metadata-Driven AI stays constant: the ratio $\mathcal{T}_{ci} / \mathcal{T}_{mf}$ is approximately $R$. If the bird-strike ranking is recomputed on each year's file for ten years, the agent pays ten sessions and the pattern one. In the bank-statement case study (Section~\ref{sec:pdf}), $R$ is the number of statements of a given layout processed over the lifetime of the extractor, which can reach hundreds or thousands per year.

The architectures fetch the same external data at run time, so network transfer is not where they differ. Translating tokens into energy requires per-token energy estimates, which vary widely with models and hardware \cite{luccioni2024power}; we therefore report token counts, which can be measured directly, and leave the conversion to energy to the reader's assumptions.

\subsection{What Crosses the Inference Boundary}

To quantify the difference between $T_{meta}$ and $T_{data}$ on tabular data, we applied a metadata extractor we built to four public datasets from the vega-datasets collection \cite{vegadatasets}. For each column, the extractor emits its type and range (numbers), the full list of values (categorical columns with at most 50 distinct values), or the number of distinct values and three examples (other strings). Table~\ref{tab:metadata_size} compares the size of the raw CSV file with the size of the metadata that is actually sent to the LLM.

\begin{table}[t]
\centering
\caption{Size of the raw data vs.\ size of the metadata sent to the LLM for four public datasets derived from the vega-datasets collection \cite{vegadatasets}. Metadata as produced by our extractor (column types, ranges, category lists). One token is roughly four bytes of English text.}
\label{tab:metadata_size}
\begin{tabular}{@{} l r r r r r @{}}
\toprule
\textbf{Dataset} & \textbf{Rows} & \textbf{Columns} & \textbf{Raw data (bytes)} & \textbf{Metadata (bytes)} & \textbf{Ratio} \\
\midrule
birdstrikes      & 10,000 & 14 & 1,213,329 & 4,018 & 302$\times$ \\
seattle-weather  &  1,461 &  6 &    48,219 &   361 & 134$\times$ \\
disasters        &    803 &  3 &    18,840 &   318 &  59$\times$ \\
gapminder-health-income & 187 & 5 & 8,605 &   409 &  21$\times$ \\
\bottomrule
\end{tabular}
\end{table}

The metadata is between one and three orders of magnitude smaller than the data, and the ratio grows with the number of rows, since the metadata of a column is bounded (a range, at most 50 categories, or three examples) whereas the data grows linearly. The 1.2~MB birdstrikes table of Section~\ref{sec:example} (roughly 300,000 tokens) would exceed the context window of most small and locally hosted models, and fill a large part of that of the largest frontier models, if placed in context, while its metadata fits in about 1,000 tokens. A code-interpreter agent would not place the whole table in context either, but it would upload it to the provider and typically inspect a preview of its rows, which, unlike the metadata above, is not bounded by design. Note that metadata is not devoid of information: ranges and category lists are real values. We discuss this point in Section~\ref{sec:leakage}.

\subsection{Can a Small Local Model Be the Generator?}
\label{sec:smallmodels}

Sovereignty is only complete if the Generator itself can run on infrastructure the organisation controls. Because the Generator writes code from metadata rather than reasoning over the data, the task is closer to code completion than to open-ended analysis. We built a small benchmark that grades generated formulas mechanically, by executing them and comparing their output with golden values computed from the data: an aggregation task (the produced series must sum to the dataset total within 0.5\%), a regression task (the fitted trend must be finite and bounded, which catches fabricated coefficients), and a refinement task (a follow-up request must produce a complete, runnable formula). Table~\ref{tab:bench} reports the results.

\begin{table}[t]
\centering
\caption{Pass rates of generated formulas on our benchmark (disasters dataset, 3 runs per task). Latency is the median time of one generation round-trip.}
\label{tab:bench}
\small
\begin{tabular}{@{} l l l r r r r @{}}
\toprule
\textbf{Model} & \textbf{Thinking} & \textbf{Runs on} & \textbf{Aggr.} & \textbf{Regr.} & \textbf{Refin.} & \textbf{Latency} \\
\midrule
gemini-3-flash-preview & low     & cloud & 100\% & 100\% & 100\% & 7.4 s \\
gemini-3-flash-preview & default & cloud & 100\% &  67\% & 100\% & 10.5 s \\
gemini-3.5-flash       & default & cloud &  67\% &  33\% & 100\% & 32.3 s \\
qwen3:8b \cite{qwen3}  & low     & laptop$^{*}$ & 100\% & 100\% & 100\% & 22.6 s \\
\bottomrule
\end{tabular}

\smallskip
{\footnotesize $^{*}$Apple M5, 16 GB RAM, via Ollama.}
\end{table}

An 8-billion-parameter open-weight model running on a laptop passed all tasks, at a latency that is acceptable for a one-time generation step. These results are indicative only (three tasks, three runs, one dataset), but they support the claim that, in this pattern, the Generator does not need to be a frontier cloud model.

\section{Case Studies}
\label{sec:cases}

We now show the pattern at work on three applications from different domains, each instantiating it differently: a geospatial prospection pipeline over live public APIs, the extraction of portfolio positions from bank statements, and a conversational assistant over smart-city dashboards. For each, we give the context, how the pattern is instantiated, the generated formula, and what we obtained; a precise quantitative evaluation is left for future work.

\subsection{Geospatial Photovoltaic Prospection}
\label{sec:geo}

The first example applies the pattern to an analysis that joins public, frequently updated sources: the identification of companies whose roofs and parking lots could host photovoltaic (PV) panels in France. It illustrates the \textit{just-in-time} variant, in which the data is not held locally but fetched through the API Proxy at run time.

\paragraph{Phase 1: target identification} The user asks for energy-intensive companies within a given area. The Generator writes a formula that queries the French government's open company register API (\url{https://recherche-entreprises.api.gouv.fr}), filters companies by activity codes and financial indicators (turnover, number of employees), and returns a layer of prospects. The Generator only knows the API's contract, not its content.

\begin{figure}
    \centering
    \includegraphics[width=0.75\linewidth]{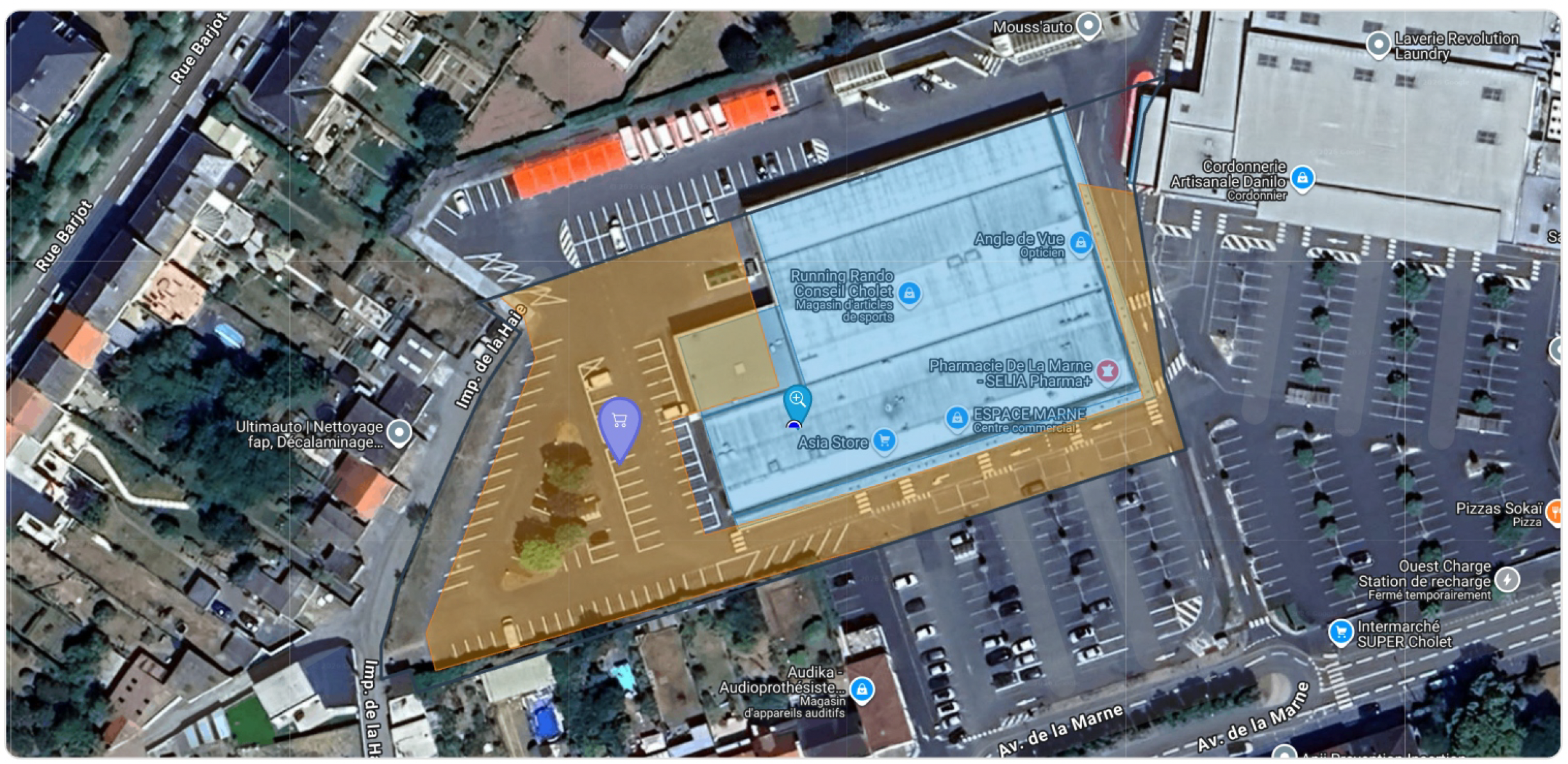}
    \caption{Result of the solar audit formula for one prospect: building footprints (blue) and parking lots (orange) intersected with the cadastral parcel, with the computed areas added to the prospect's data.}
    \label{fig:screenshot}
\end{figure}

\paragraph{Phase 2: solar potential audit} The user then asks for the surface area available for PV panels on a given prospect. The Generator writes a formula (Listing~\ref{lst:geo}) that fetches the building footprints from the French national mapping agency's WFS service (IGN BD TOPO), the parking lots from OpenStreetMap via the Overpass API, intersects them with the cadastral parcel of the company using \texttt{turf.js}, computes the areas, and draws them on a map (Figure~\ref{fig:screenshot}). All of this runs in the sandbox; the API calls go through the API Proxy, which only allows the listed endpoints.

\begin{lstlisting}[language=JavaScript, label={lst:geo}, caption={Excerpt of the generated solar audit formula. External GIS and REST services are called just-in-time through the API Proxy, and the geometry is computed locally.}]
// 1. Fetch Buildings (IGN BDTOPO via WFS)
const urlBat = `https://data.geopf.fr/wfs/ows?SERVICE=WFS&VERSION=2.0.0` +
               `&REQUEST=GetFeature&TYPENAMES=BDTOPO_V3:batiment` +
               `&BBOX=${bbox.join(',')},urn:ogc:def:crs:EPSG::4326&OUTPUTFORMAT=application/json`;
const resBat = await fetch(urlBat).then(r => r.json());

// 2. Fetch Parking Lots (Overpass API)
const opQuery = `[out:json][timeout:25];(way["amenity"="parking"]` +
                `(${bbox[1]},${bbox[0]},${bbox[3]},${bbox[2]}););out body;>;out skel qt;`;
const resOp = await fetch(`https://overpass-api.de/api/interpreter?data=` +
                           encodeURIComponent(opQuery)).then(r => r.json());
// 3. Local Execution: Calculate Intersection Surface Area
let totalBuildingArea = 0;
if (resBat.features) {
    for (const f of resBat.features) {
        const inter = turf.intersect(parcelGeom, f); // resolved locally via turf.js
        if (inter) {
            totalBuildingArea += turf.area(inter);
            drawToMap(inter, { fillColor: '#3498db' });
        }
    }
}
\end{lstlisting}

Once validated on one prospect, the audit formula is applied to the whole prospect layer at run time: it iterates over hundreds of companies, fetches the geometries for each, and writes the computed areas back into the local data, without any LLM call.

\paragraph{Lessons learned}

The value of the pattern here shows at run time. The audit formula is generated once, from the API contracts alone, and then applied to the whole prospect layer --- hundreds of companies, each needing several API calls and local geometry --- without a single further LLM call. Inference is thus paid once for the whole campaign instead of once per prospect or per run, so the saving in tokens and cost grows with the number of prospects and with how often the audit is repeated. The pattern also keeps data access live: building footprints, parking lots, and the company register are queried just-in-time through the API Proxy, so no national dataset has to be replicated and kept in sync locally. Its main liability is the reliance on these external APIs, whose changes can break a stored formula (Section~\ref{sec:apivolatility}).

This case study is deployed in production as a commercial Salesforce plugin on the AppExchange, built on its own implementation of the pattern --- in itself evidence that the pattern can be realised independently of any particular codebase.

\subsection{Extracting Portfolio Positions from Bank Statements}
\label{sec:pdf}

The second example applies the pattern to semi-structured documents. It was developed for a wealth-management customer who needs to fill portfolio information (accounts, securities, quantities, prices, valuations) from the statements sent by custodian banks, as PDF documents or spreadsheet exports; a single statement may cover several accounts. Each bank uses its own layout, but a given bank's statements keep the same layout from one month to the next. The task is therefore highly repetitive within a layout and heterogeneous across layouts, which is precisely the situation the pattern addresses.

\paragraph{From one sample to a reusable extractor} One sample statement is enough to create an extractor. An administrator loads a statement of a given bank and document type and asks for its accounts and positions. The Generator writes an extraction formula, which is run on the sample at once; the administrator points out what is wrong in plain language, and the Generator amends the formula, typically within two to four requests for a difficult multi-account statement. The formula is then saved as the extractor of that bank and document type, and every later statement of that type is processed by it alone, in the browser and without any AI, until the bank changes its layout. Adding a bank costs one design-time session; processing a new month's statements costs no inference at all. The platform previously sent every statement through optical character recognition and an LLM; that path now remains only as a fallback for banks that have no extractor yet. This is the programming-by-example idea \cite{gulwani2011automating} applied with a general-purpose model: one example suffices because the Generator is asked to separate the \textit{static} parts of the document (column headers, section captions, field labels) from its \textit{dynamic} parts (positions, amounts, dates, number of lines), and to locate the latter only through the former.

\paragraph{Metadata: a structure-preserving exemplar} For documents, the metadata is a text rendering of the sample that preserves its layout. It is computed deterministically, without the LLM, by an edge service of the Orchestrator that runs on a server of the organisation. PDF pages are rendered with \texttt{pdfplumber} \cite{pdfplumber}: column boundaries come from the ruling lines of the page or, on borderless pages, from the horizontal gaps between words, and indentation is kept so that the hierarchy of categories and line items remains visible. A spreadsheet is rendered sheet by sheet. Columns are separated by a \texttt{|} character and pages by a header (anonymised example):
\begin{lstlisting}[numbers=none]
## Page 2
Compte | 0000000.001 | Compte titres | EUR
Code valeur | Libellé | Quantité | Cours | Valorisation
  XX0000000001 | EXAMPLE EQUITY FUND A | 1 250 | 102,35 | 127 937,50 €
\end{lstlisting}

The same rendering is computed for every future statement and passed to the formula at run time. When the rendering was wrong (merged cells, truncated text), asking the Generator to work around it failed, whereas a fix in the rendering corrected every document at once: structural problems are best fixed before the inference boundary, not in the prompt.

\paragraph{Inference boundary and feedback} Here the metadata is a whole sample document, so its content does cross the boundary at design time: the structure of a document cannot be described without showing it. Exposure can be limited by the choice of the sample, by masking its dynamic values, or by using a local Generator; the customer chose to create its extractors from non-critical statements of its own. The statements processed at run time, which are the vast majority, never reach the model. During authoring, the Orchestrator adds an execution trace to each new request: the number of accounts and positions extracted, and the positions grouped by the fields they fill, with a few examples. This is the refinement loop with filtered feedback, since the trace only carries values of the sample, which the Generator has already seen. The trace reports what was extracted, never whether it is correct: we found that a trace declaring the extraction complete competed with the user's feedback, and that the model tended to side with the trace. The Generator is a frontier cloud model (Gemini~3~Pro). With the same instructions, a smaller model of the same family repeatedly selected rows by keywords in their labels instead of by their structure: documents with complex layouts may need a larger Generator than the tabular tasks of Section~\ref{sec:smallmodels}, which remains affordable since it is used once per layout.

\paragraph{Structure of a generated extractor} The instructions ask for a formula (Listing~\ref{lst:pdf}) that returns the statement date and its accounts, each with its positions; that first checks the static anchors of the layout and stops with an explicit error when they are missing; that classifies every line before extracting values from the lines classified as positions; that tells positions from subtotals by their structure (an ISIN on the row or on the adjacent line, the indentation), never by keywords, since an instrument can be named ``TotalEnergies''; that parses numbers with the provided helpers, which handle European and Anglo-Saxon formats; and that hard-codes no value. These instructions are kept short and stated as principles: longer lists of rules degraded the extraction of ordinary statements.

\begin{lstlisting}[language=JavaScript, label={lst:pdf}, caption={Illustrative extraction formula (anonymised, simplified). \texttt{pdfText} is the structure-preserving rendering of the statement; \texttt{splitCols}, \texttt{parseNum} and \texttt{parseMoney} are helpers provided by the Executor. The formula returns the statement date and its accounts, each with its positions.}]
// Structure check: static anchors of this statement layout
if (!/Code valeur \| Libell[eé] \| Quantit[eé]/i.test(pdfText))
  return { html: '<p>Unexpected document structure.</p>',
           data: { error: 'structure check failed' } };
const ISIN = /^[A-Z]{2}[A-Z0-9]{9}[0-9]$/;
const lines = pdfText.split('\n');

// Pass 1: classify every line by its structure
const kinds = lines.map(l => {
  const c = splitCols(l);
  if (/^Compte$/i.test(c[0]) && c.length >= 4) return 'account';
  if (ISIN.test(c[0]) && c.length >= 5) return 'position';
  return 'other';            // headers, subtotals, free text
});

// Pass 2: extract values from the classified lines
const accounts = []; let acc = null;
lines.forEach((l, i) => {
  const c = splitCols(l);
  if (kinds[i] === 'account')
    accounts.push(acc = { account_number: c[1], account_type: c[2],
                          currency: c[3], positions: [] });
  if (kinds[i] === 'position' && acc)
    acc.positions.push({ isin: c[0], name: c[1],
      quantity: parseNum(c[2]), price: parseMoney(c[3]),
      market_value: parseMoney(c[c.length - 1]) });
});
const d = pdfText.match(/Situation au (\d\d)\/(\d\d)\/(\d{4})/);
return { html: renderAccounts(accounts),   // HTML tables, omitted
         data: { statement_date: d ? d[3] + '-' + d[2] + '-' + d[1] : null,
                 accounts } };
\end{lstlisting}

Extraction is followed by checks that do not involve the model: a deterministic coherence check flags positions whose value disagrees with other estimates, such as quantity times price, and a human review compares, for each account, the sum of the positions with the total printed on the statement before anything enters the portfolio. Such checks turn silent extraction errors into visible ones, which matters more than average accuracy here. A quantitative evaluation (layouts and statements processed, precision and recall, onboarding time versus manual entry) is left for future work.

\subsection{Conversational Analysis of City Dashboards}
\label{sec:city}

The third example is the Clima Platform, the city platform of the Horizon Europe project CLIMABOROUGH (2023--2026) \cite{climaborough}, in which eight European cities test climate solutions in real conditions and monitor their impact towards climate neutrality. LIST builds and operates the platform's dashboards on BESSER, its open-source low-code platform \cite{alfonso2024besser}: the domain model is written in B-UML, BESSER's modelling language based on UML \cite{besserdocs}, the database, REST API and front-end services are generated from it, and city officers compose dashboards of charts without programming \cite{conrardy2025lowcode}. The data is a set of key performance indicators (KPIs) per city, such as waste fractions, recycling rates or energy use, each a monthly or hourly time series, optionally split by category. The platform is publicly accessible \cite{climaplatform}, and its AI assistants use their own implementation of the pattern.

\paragraph{Instantiation of the pattern}

The participants of the pattern map onto the platform as follows. The Orchestrator is the assistant service, which runs next to the back end. The browser sends it only the state of the chart in view, never values; the Orchestrator reads KPI definitions and values itself from the platform's API. The Generator can be any LLM served through an OpenAI-compatible endpoint, from a cloud model to a self-hosted one. The Executor is the set of generated chart components of the front end, which render the widgets the Generator creates, and, for chart explanations, the Orchestrator's own statistics code. The Formula Repository is the platform's database, where widgets are stored with their dashboard. No API Proxy is needed, since formulas only read the platform's own data. The metadata comes from the domain model. Besides its name, unit and thresholds, each KPI carries optional interpretation attributes that we added to the B-UML model: its aggregation type (e.g., share or cumulative), its direction (higher or lower is better), whether its categories form a composition, a baseline, a target, and a disclosure level. They tell the Generator how to read a statistic.

\paragraph{Generated formulas as model instances}

In edit mode, the assistant creates charts from requests such as \emph{show the residual waste share over time}. The Generator receives the city's KPI catalogue (identifier, name, unit and thresholds of each KPI) and no values, and returns a widget specification (Listing~\ref{lst:clima-widget}). Here the formula is not code but a declarative specification, an instance of the B-UML model: the Orchestrator checks that its type is one of the model's chart types and that its KPI is in the catalogue. The front end completes it with layout defaults and with the KPI's unit and thresholds, and adds it to the dashboard; when the dashboard is saved, the widget is stored with the others. Each time the dashboard is opened, the chart component that matches the widget's type reads the specification, fetches the KPI's current values from the platform's API, aggregates them by period and draws the chart, without calling the LLM. This is the declarative form of the formula described in Section~\ref{sec:pattern} (\textit{Variants}): the generated formula is not a program but a declarative model, used by an interpreter or a function that plays the Executor role, here the chart component. This is more restrictive than executing a program, but it is much safer, and it avoids having to implement a sandbox that restricts what the program can do and ensures that LLM-generated code does not execute malicious operations.

\begin{lstlisting}[numbers=none, label={lst:clima-widget}, caption={A formula in the city-dashboard case: the widget specification the Generator returns for ``show the residual waste share over time'', given the KPI catalogue only. Once saved, it is rendered on new data without inference.}]
{ "answer": "I added a time chart of the residual waste share.",
  "visualization": { "type": "TimeChart",
                     "kpi_id": 3,
                     "title": "Residual waste share" } }
\end{lstlisting}

\paragraph{A variant: explaining charts}

\begin{figure}
    \centering
    \includegraphics[width=0.9\linewidth]{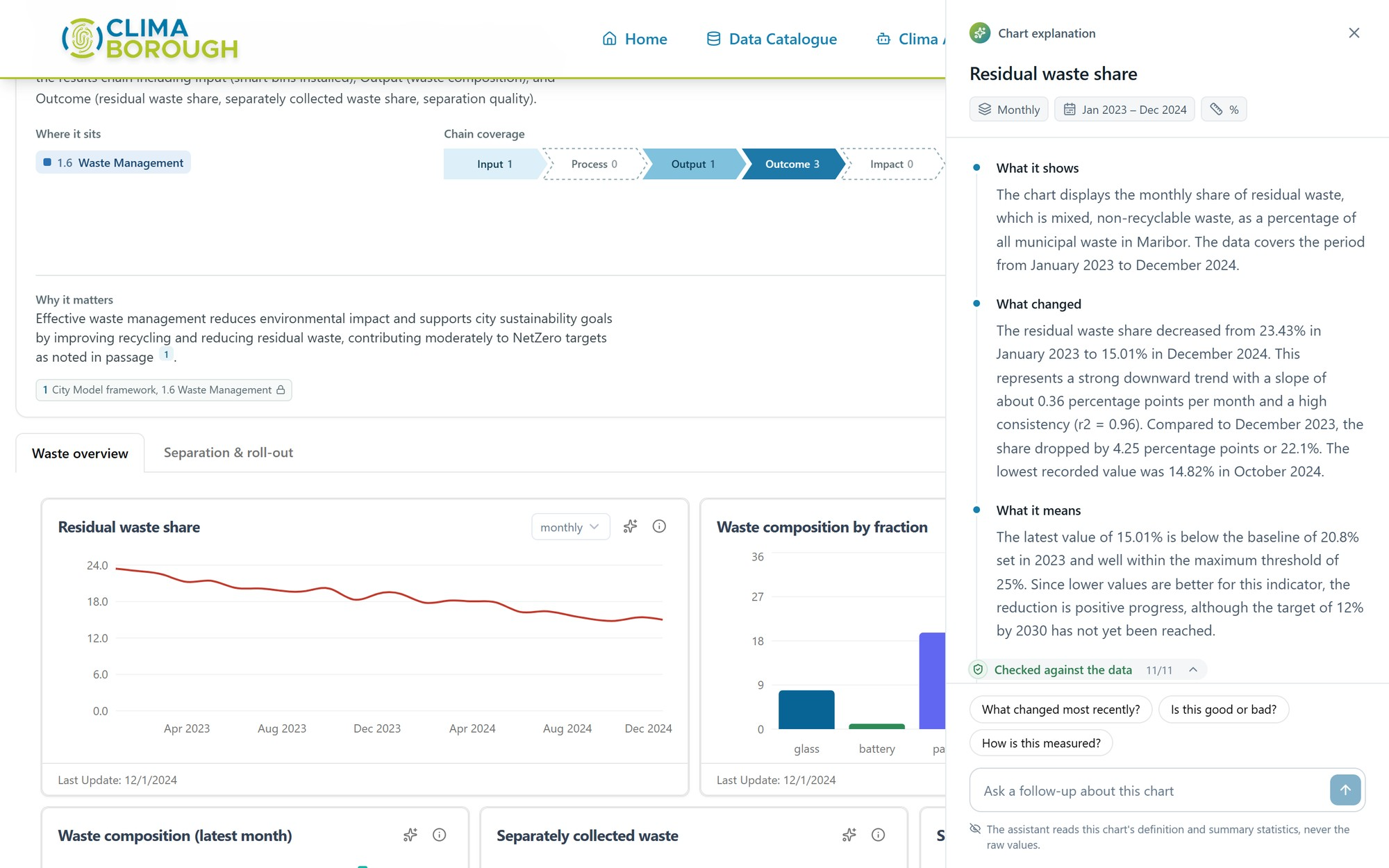}
    \caption{Explaining the \emph{Residual waste share} chart of a Clima Platform dashboard (local Maribor demo data). The badge reports that all 11 figures in the answer match the statistics computed locally.}
    \label{fig:clima}
\end{figure}

In view mode, every chart can explain itself (Figure~\ref{fig:clima}). This assistant departs from the pattern on two points. The formula is not generated: it is a profile function written once by the developers, which the Orchestrator runs locally on the KPI's values to compute summary statistics (latest value, changes, trend, extremes, distance to the target). And its result crosses the boundary: the Generator receives the metadata and this profile (Listing~\ref{lst:clima-bundle}), never the series, and puts the profile into words. The Orchestrator limits what crosses in three ways. The profile has a fixed maximum size, whatever the number of rows. The KPI's disclosure level decides what it keeps, from summary values at \texttt{public} to directions and shares only at \texttt{restricted}. And a verifier matches every number of the answer against the numbers that crossed; if one does not match, the answer is regenerated once, and sentences that still contain unmatched numbers are removed.

\begin{lstlisting}[numbers=none, label={lst:clima-bundle}, caption={What crosses the inference boundary to explain the monthly \emph{Residual waste share} chart (excerpt; local Maribor demo data, 24 monthly values). Interpretation attributes come from the B-UML model; the profile is computed locally. The series itself is not sent.}]
"kpi": { "name": "Residual waste share", "unit": "%",
         "max_threshold": 25.0, "aggregation_type": "share",
         "direction": "lower_is_better",
         "baseline": {"value": 20.8, "year": 2023},
         "target": {"value": 12.0, "year": 2030},
         "disclosure_level": "public" },
"profile": { "series": [{ "n": 24,
    "first":  {"period": "2023-01", "value": 23.43},
    "latest": {"period": "2024-12", "value": 15.01},
    "change_prev": {"abs": -0.39, "pct": -2.5, "vs": "2024-11"},
    "change_yoy":  {"abs": -4.25, "pct": -22.1, "vs": "2023-12"},
    "trend": {"direction": "down", "slope_per_period": -0.3565,
              "r2": 0.96, "strength": "strong"},
    "min": {"period": "2024-10", "value": 14.82},
    "vs_baseline": {"abs": -5.79, "pct": -27.8} }] }
\end{lstlisting}

\paragraph{What we obtained}

The answer in Figure~\ref{fig:clima} states, for instance, that \emph{``the residual waste share decreased from 23.43\% in January 2023 to 15.01\% in December 2024''} and that, \emph{``compared to December 2023, the share dropped by 4.25 percentage points or 22.1\%''}. All eleven numbers in the answer are fields of Listing~\ref{lst:clima-bundle}, and the verifier matched all of them, so the answer took a single LLM call and no regeneration. When we replaced the October minimum (14.82\%) with 13.9\% in an answer, the verifier flagged it and the sentence was removed. The prompt held 1,597 tokens, 461 of them for the bundle.

\paragraph{Lessons learned}

In a model-driven platform, the domain model provides both the metadata and the language of formulas: a chart created by the assistant is validated, stored and re-run like a chart built by hand. Chart creation follows the pattern fully. Chart explanations answer the fixed set of questions that the profile covers, and they need a governed result to cross the boundary. Letting the Generator write new profile functions on demand, stored and checked like the existing one, would bring them within the pattern. Two limits of Section~\ref{sec:leakage} also apply here: at the \texttt{public} level, the latest value and the extremes of a series are individual observations; and the verifier only checks that a number appears in the profile, not that it describes the right quantity.

\section{Discussion and Limitations}
\label{sec:discussion}

\subsection{Comparison of Case Studies}

We have successfully instantiated the pattern in three very different contexts --- live geospatial APIs, PDF bank-statement extraction, and a smart-city dashboard assistant. This is not a formal evaluation, but it is a first indication that the pattern is complete enough to cover a broad range of data-analysis situations. Table~\ref{tab:cases} summarises how the pattern was instantiated in each: the same participants appear every time, while the nature of the metadata, of the reuse, and of the optional participants needed differs. The three were, moreover, built with three different implementations of the pattern --- a commercial Salesforce plugin, our own implementation, and an open-source BESSER-based assistant --- so its generality does not rest on any single codebase.

\begin{table*}[t]
\centering
\caption{Instantiation of the pattern in the three case studies.}
\label{tab:cases}
\small
\renewcommand{\arraystretch}{1.25}
\begin{tabularx}{\textwidth}{@{} p{2.8cm} >{\raggedright\arraybackslash}X >{\raggedright\arraybackslash}X >{\raggedright\arraybackslash}X @{}}
\toprule
 & \textbf{Geospatial prospection} & \textbf{Bank statements} & \textbf{City dashboards} \\
\midrule
Data & live public APIs (company register, IGN, OpenStreetMap) & PDF and spreadsheet statements & Dashboard indicator \\ \addlinespace
Metadata & API contracts & structure-preserving text of one sample & Schemas and statistics, or the B-UML model \\ \addlinespace
Crosses the boundary & no data & one sample document (can be masked) & no data \\ \addlinespace
Formula & API orchestration and geometry & line classification and extraction of accounts and positions & declarative widget specification (model instance) \\ \addlinespace
Reuse & new prospects and areas & every future statement of the same layout & widgets re-rendered on new values \\ \addlinespace
API Proxy & required: allow-list, credentials, just-in-time fetching & not needed & not needed \\ \addlinespace
Edge services of the Orchestrator & none & text extraction of PDF, spreadsheet and Word documents, on a server of the organisation & none \\
\bottomrule
\end{tabularx}
\end{table*}

\subsection{When Is Metadata Enough?}
\label{sec:enough}

In building and using the pattern, we observed that LLMs are often strikingly good at writing correct code from metadata alone: for many analyses a schema with a few statistics is enough for the Generator to produce a formula that computes the right answer, and showing the model the data itself would add exposure without improving --- and occasionally while degrading --- the result. This runs against the intuition that a model reasons better the more of the data it sees. We report it as an observation from our experience rather than a measured finding: we have not yet evaluated \textit{when} code generation from metadata outperforms direct inference over the data, nor \textit{which} forms of metadata extraction (which statistics, which exemplars, which structural summaries) most help the Generator write accurate code. Characterising this trade-off, and the preprocessing that best supports it, is left for future work.

\subsection{Metadata Is Not Zero Information}
\label{sec:leakage}

Keeping data out of the model's context does not mean that no information leaves. Value ranges reveal minima and maxima; category lists reveal the values of low-cardinality columns; examples of string values are real values; and, for small tables, the category list may contain the entire column. For documents, the sample itself crosses the boundary. The pattern therefore reduces exposure by design and makes it auditable (the Orchestrator can log every request), but it does not make it zero. A second channel is \textit{feedback}. To repair a formula, it is useful to return its error messages or logs to the Generator, and these logs may contain values from the data; formulas that parse documents are even asked to log what they read at each step, precisely so that problems can be diagnosed. This is the same channel that code-interpreter agents use by default and without restriction. In the pattern, it is closed by default and opened only under the Orchestrator's control, which the application configures and which records every message. How much is exposed therefore depends on how the application instantiates the pattern. Since the application knows which data is sensitive, the Orchestrator can redact these values before they are sent, flag suspicious content, or show the feedback to the user for approval. A third channel is the user, who may paste data into the intent. The Orchestrator can close it cheaply, because it holds the data: it matches the prompt against the data and refuses to send a prompt that contains some of it. Trusted research environments face the same issue and address it with output checking before release \cite{williamson2020opensafely}, which is the role the Orchestrator plays here. Our implementation keeps runtime feedback off unless the host application enables it, but does not yet implement redaction; automated redaction of formula logs is a direction for future work. Implementations should let users choose the granularity of metadata for each dataset (schema only, schema with statistics, or masked statistics), and sensitive deployments should use a Generator running on infrastructure they control, which Section~\ref{sec:smallmodels} suggests is feasible.

\subsection{API Volatility}
\label{sec:apivolatility}

A formula that calls external APIs breaks when those APIs change. This is frequent: in an empirical study of 2,224 versions of RESTful APIs, Yasmin et al.\ found that 87.3\% of the APIs that introduced breaking changes did not deprecate the affected operations beforehand \cite{yasmin2020first}. The same holds for document layouts, which banks change without notice. The pattern mitigates this in two ways. Formulas fail fast and explicitly (structure checks, logged errors) instead of returning wrong results. And since the original intent and the error are both available, the Generator can repair the formula in a short design-time session, after which the repaired formula is again reusable.

\subsection{Evolution of the Data}
\label{sec:evolution}

A stored formula encodes assumptions about the data it was generated for, and the data may change. Some changes are structural --- a column is renamed or removed, a section of a document moves --- and the declared dependencies catch them: the dependency check, or the formula's own structure check, stops the run before a wrong result is produced. Other changes alter the meaning or the domain of the values without changing the structure: a cost column is now reported in thousands of dollars, a categorical column gains a value that should have been grouped with an existing one, values now fall outside the intervals a formula uses to bin them, or a status code is redefined. The formula still runs and returns a plausible but wrong result, and nothing in that result signals that the formula is no longer aligned with the data. This is the price of reuse: an approach that regenerates the analysis at every run adapts to such changes implicitly, at the cost of the reproducibility that the pattern provides, whereas a stored formula keeps doing what it did.

Two consequences follow. First, the pattern places some responsibility on the user, who must understand why a formula that used to work now fails, or why its results look odd, and know that a new design-time session is needed to bring the formula back in line with the data. Users who cannot read code depend on the explanation that accompanies the formula and on the quality of its error messages. Second, the Orchestrator can take on part of this responsibility, because it can keep the metadata a formula was generated from: at run time, it can derive the metadata of the new data, compare the two, and warn the user, or propose regeneration, when types, value ranges, or category lists have changed. The output contract can help too, by asking the Generator to turn the assumptions a formula relies on (a unit, a range, a closed list of categories) into explicit checks rather than leaving them implicit. Neither measure catches every change --- a status code redefined with the same set of values would go unnoticed --- and we have not yet evaluated such drift detection; it is a direction for future work.

\subsection{Client-Side Resources}

When the Executor is a browser sandbox, formulas are bounded by the memory and the mostly single-threaded execution of the client. Simple aggregations, parsing, and geometric operations run comfortably on thousands of items, but heavy computations (large spatial joins, model training) are better served by a server-side Executor, which the pattern allows. Large workloads can also be split into batches whose progress is persisted locally. The gap between servers and end-user devices is narrowing \cite{siffre2025local}, and WebAssembly makes lower-level, high-performance code available in the browser. The pattern is best suited to the enrichment and transformation of distributed data and to repetitive extraction; it does not aim to replace databases for global queries over large static datasets.

\subsection{Governance of the Formula Repository}

When examples are retrieved from the Formula Repository, stored formulas improve generations only if they are good. Poor or erroneous formulas in a shared repository would be retrieved and propagated, and RAG systems are known to be vulnerable to corrupted knowledge: PoisonedRAG showed that injecting a few crafted texts into a large knowledge base was enough to control the output of the model for targeted questions with high success rates \cite{zou2025poisonedrag}, and similar attacks extend to multimodal RAG \cite{ha2025mm}. Even without malicious intent, a repository needs curation: formulas should be validated before being shared (ideally with mechanical checks similar to the benchmark of Section~\ref{sec:smallmodels}), versioned, and associated with the structure they apply to.

\subsection{Security of Generated Code}

Executing generated code on real data requires a sandbox that prevents it from exfiltrating data. In our implementation, the sandbox has no direct network access, and the only outbound channel is the API Proxy, which enforces an allow-list of APIs. The generated code is also readable before it runs, which is an advantage over opaque agents, but users cannot be expected to review every formula, so the sandbox and the API Proxy remain the primary protection. The declarative variant of the pattern (Section~\ref{sec:pattern}, \textit{Variants}) avoids the issue altogether: when the formula is a specification interpreted by a trusted function, nothing returned by the LLM is executed as code, and an application that can live with the interpreter's expressiveness gets the pattern's guarantees without a sandbox, as the city-dashboard case study does (Section~\ref{sec:city}).

\section{Conclusion}

Analysing data with an LLM usually means sending that data to the model, which brings potentially significant inference costs and a loss of control over sensitive data. Moreover, if the LLM chooses to analyse the data by inference, stochastic effects may appear. And even when the model generates code to analyse the data, that code is not reusable and may lead to different results across runs.

Metadata-Driven AI is an architectural pattern that helps application designers solve all these issues in an elegant manner. It follows a principle inherited from trusted research environments: bring the code to the data, not the data to the analyst. A one-way inference boundary enforces this principle. Only metadata and the user's intent ever reach the model by default, while other information, such as debugging feedback, can be fully controlled and filtered by the application. In addition, the generated program is a first-class, persistent artifact that can be re-run on new data of the same shape without any further inference. Inference cost thus becomes independent of both the volume of data and the number of runs, and the results are reproducible and auditable. Our experience shows that the pattern works well because LLMs can generate working code from metadata alone, even with a small model that the data owner hosts.

We described the pattern in the vocabulary of design-pattern catalogues and instantiated it in three different examples, each on a different implementation. First, a geospatial prospection Salesforce application that uses live public APIs in the generated prospection formula. Second, a bank-statement extraction application for portfolio management, which turns a given statement into an extractor --- a reusable formula. That extractor then processes every future statement of the same layout with no AI, avoiding unreasonable costs and stochastic effects during extraction. Third, a smart-city assistant built on a low-code platform, which derives its metadata from the application's own model. In each case, the same participants recur. What varies is the form of the metadata, the optional participants, and the unit of reuse. 

Several directions follow from this work. The case studies presented here remain qualitative, so a natural next step is a systematic evaluation. We also plan to create an extended benchmark of small, locally hosted models used as Generators across more tasks and datasets. Beyond evaluation, we would like to package the pattern as a generic, multi-purpose library --- bundling the Orchestrator, the sandboxed Executor with its helper libraries, and the Formula Repository. This library will be exposed as an agent skill \cite{anthropic2025skills} or through the Model Context Protocol \cite{anthropic2024mcp} so that the model can include the pattern and drive it by itself. An assistant could then adopt the inference boundary from within an ordinary conversation, keeping the data on the user's side while still writing and reusing formulas. The guarantee would hold only if the tools the assistant calls never return the data, or the results of a formula, into the model's context: results go to the user, the model receives a status, and the sandbox has no network access of its own. In this way, the pattern becomes something a model can choose to apply, rather than only something an application designer instantiates by hand.

This last point --- a model deciding when to apply the pattern --- is itself worth studying. The ability of LLMs to generate correct code from an intent is already well benchmarked, from general program synthesis \cite{austin2021program} to text-to-SQL over a schema \cite{yu2018spider}. The open question here is a different one: not whether the model can write the code, but when writing it from metadata alone is enough, and when the task genuinely requires reading the data (Section~\ref{sec:enough}). We observed that code from metadata is often as good as reasoning over the data, and sometimes better, yet we do not know where that boundary lies, nor which forms of metadata best support accurate generation. A benchmark built around this trade-off --- rather than around code correctness in isolation --- would let a system route each request to the cheaper mode whenever it suffices, a decision that is central to frugal AI. The pattern may also generalise beyond data analysis: in a model-driven engineering setting such as BESSER, the artifact produced is itself code and the metamodel plays the role of the metadata, so the same boundary could govern code generation from a model rather than analysis over data. Finally, two of the pattern's softest edges deserve dedicated work. The first is richer techniques for extracting metadata from semi-structured documents, where the structure must be conveyed without revealing the content. The second is the masking of sensitive parameters wherever they may surface, whether in the metadata, in the feedback logs, or in the intent itself. 

\bibliographystyle{plainnat}
\bibliography{biblio}

\end{document}